\documentclass[aps,prl,superscriptaddress,10pt,onecolumn,amsmath,floatfix]{revtex4-1}

\usepackage{graphicx}
\usepackage{amsmath, amsthm, amssymb}
\usepackage{url}
\usepackage{color}

\begin{document}

\title[Geographic Dependence of the Solar Radiation Spectrum at Intermediate to High Frequencies]{Geographic Dependence of the Solar Radiation Spectrum at Intermediate to High Frequencies}
\author{G. Bel}
\affiliation{Department of Solar Energy and Environmental Physics, Blaustein Institutes for Desert Research, Ben-Gurion University of the Negev, Sede Boqer Campus, 84990, Israel}
\email{bel@bgu.ac.il}
\affiliation{Center for Nonlinear Studies (CNLS), Theoretical Division, Los Alamos National Laboratory, Los Alamos, NM, 87545, USA}

\author{M. M. Bandi}
\affiliation{Nonlinear and Non-equilibrium Physics Unit, OIST Graduate University, Onna-son, Okinawa, 904 0495, Japan.}
\email{bandi@oist.jp}

\date{\today}

\begin{abstract}
The solar radiation spectrum is important for renewable energy harvesting, as well as climate and ecological systems analysis. In addition to seasonal/annual and diurnal oscillations, the signal is composed of different frequencies ($f$) that modulate each other. This results in a continuous spectrum with a power-law dependence over the entire frequency range from 1/year up to at least 1/min. Geographical location plays an important role in determining the temporal variability of solar radiation and naturally affects its spectral power-law decay, primarily due to the latitudinal dependence of daylight duration and its seasonal variation,. Employing a model to calculate the clear sky solar radiation intensity at various global locations, we show that significant spatial variability exists, particularly versus the latitude, which we quantify via the exponent of the spectral power-law. At most locations, the spectral power-law dependence of the intermediate (1/day $< f <$ 1/hour) frequencies is different from that of the high ($f > 1/$hour) frequencies. We demonstrate the origin of this power-law dependence from simple arguments, explain its geographic dependence and discuss the implications for photovoltaic power generation and solar-driven systems, and hence the impact on grid stability.
\end{abstract}

%

%
%
\maketitle
%
%

\section{Introduction}
Solar radiation is directly or indirectly responsible for almost all life and energy processes on the earth. It plays an important role in the analysis and modeling of climate \cite{Alpert2005, Lean1998} and ecological systems \cite{Austin2006, Bothwell1994, Stoepler2013}. Usually only slow variation in solar radiation is deemed important, and fluctuations at sub-daily timescales are averaged out, although cases to the contrary do exist \cite{Ehleringer1980, Avissar1991,Chen1999,Li2013}. In solar energy harvesting and, in particular, for solar photovoltaic power production, the fast fluctuations at frequencies $f > 1/$day are of greater importance \cite{Klima2018} because they fall in the range bounded between grid response and consumer load variation timescales. Since solar radiation intensity varies with location on the earth, particularly due to changes in the angle of incidence and diurnal duration dependence on the earth's latitude due to its obliquity (axial tilt relative to the orbital plane), it impacts all the abovementioned processes. Therefore, the importance of understanding the geographic dependence of solar radiation spectral characteristics cannot be overemphasized.

In this letter, we study the geographic dependence of the solar radiation spectrum at various locations on the earth using data generated from an open source Ineichen-Perez Clear Sky Model (CSM) \cite{Perez2002, Ineichen2002}, which we first compare against a measured spectrum for a specific location (Sede Boqer, Israel, coordinates: 30.852310, 34.780586, altitude: 490 m above sea level). Analysis of spectra from real measurements is complicated by particularities such as local topographic features, strong fluctuations from cloud passage that act as multiplicative noise, etc. Calculated CSM spectra help separate these dependencies and allow one to focus on spectral features that are dependent on the geographic location alone.

Primary conclusions from our analysis of calculated spectra include the expected spectral peaks from a diurnal oscillation. Additionally, the modulation of various frequencies results in a continuous spectrum with a power-law dependence on the frequency ($f$), which has two separate slopes for intermediate (1/day $ < f < $1/hour) and high ($f > 1/$hour) frequencies. We explain the origin of the spectral power-law decay and use its exponent to quantify the geographic dependence of the spectral characteristics. As expected, the power-law exponent varies significantly with latitude for both intermediate and high frequencies, but not so much with the longitude. Finally, through a minimal toy model, we show that the slow seasonal variation of daylight duration is sufficient to explain the spectral power-law at intermediate and high frequencies. The fact that seasonal variation in daylight duration is strongly location-dependent automatically explains the strong geographic dependence of the spectral power-law exponent. We close this letter by discussing the implications of these results in the specific context of solar photovoltaic power spectra.

\section{Methods}
Solar radiation comprises three different components \cite{Spitters1986,Iqbal2012,Elsebaii2010,Li2015}: Direct (beam) radiation is the intensity of radiation traveling in a straight line from the sun to the measuring device. Diffuse radiation is the intensity of radiation scattered by clouds, local topographic features, etc. and is isotropic. Global radiation is direct radiation multiplied by the cosine of the angle between the measuring plane (assumed to be parallel to the earth's surface) and the sun plus the diffuse radiation component. The global radiation is therefore not independent of the other two. We focus on the global and direct radiation components in this study.

The solar radiation incident on the earth exhibits modulation over timescales far greater than the diurnal oscillation timescale ($f = 1/$day), from seasonal/annual periods extending to hundreds of millennia (Milankovitch cycles \cite{Petit1999, Lisiecki2005}). Therefore the solar insolation time series does not approach stationarity over any reasonable measurement time period, thereby invalidating the application of time-domain methods. Consequently, strictly spectral methods are needed. Hence our analysis is based on the power spectrum of solar radiation. The power spectrum presented here has dimensions of the square of the Fourier transform of the temporal radiation intensity, namely, units of $W^2m^{-4}{min}^2$. The Fourier transform, or more precisely series, is defined for the discrete signals that we analyze as:
\begin{align}
	\tilde{R}\left(f_n\right)&=\sum\displaylimits_{k=1}^N W_N^{kn}R\left(t_k\right);\\
	W_N&\equiv\exp\left(-2\pi i/N\right). \nonumber
\end{align}
Here $N$ is the length of the time series, and the discrete frequencies are given by $f_n\equiv n/(N\Delta t)$ with $n\in[0,N/2]$ and $\Delta t$ is the time interval. We use a year-long time series with a 1-min resolution for both measured and calculated radiation. The calculated radiation is based on CSM \cite{Ineichen2002,Perez2002}, which calculates the radiation based on the time, location, altitude and, if required, also by the air turbidity using empirical values for the location-dependent Linke turbidity index \cite{Remund2003, soda2012}, which varies monthly (according to the calendar month). A report on clear sky models found the CSM to have excellent performance \cite{Reno2012}. Calculations employing CSM with no turbidity (Linke Turbidity TL = 1) are referred to as CSMTL1.

In what follows, we extensively rely on the power-law fit of obtained spectra. By this, we mean that if the power spectral density $PS(f)$ varies with frequency $f$ with the functional form $PS(f) = A f^{\zeta}$, we say the power spectrum exhibits a power-law with an exponent $\zeta$, and the prefactor $A$ captures the magnitude of the spectrum (subtracting the mean from the original signal ensures that for the zero frequency, the power spectrum vanishes). If one takes the logarithm of both sides of the above equation, the spectrum can then be represented as $\text{log}_{10}(PS(f)) = \text{log}_{10}(A f^{\zeta}) \equiv 
\text{log}_{10}A  + \zeta \text{log}_{10}PS(f)$. In the following, we plot $\text{log}_{10}PS(f)$ versus $\text{log}_{10}{f}$ when presenting spectra so that the slope of the plot yields the power-law exponent $\zeta$, and $A$ gives us the spectral magnitude.
\begin{figure}[!ht]
  \centering
  \includegraphics[width=\columnwidth]{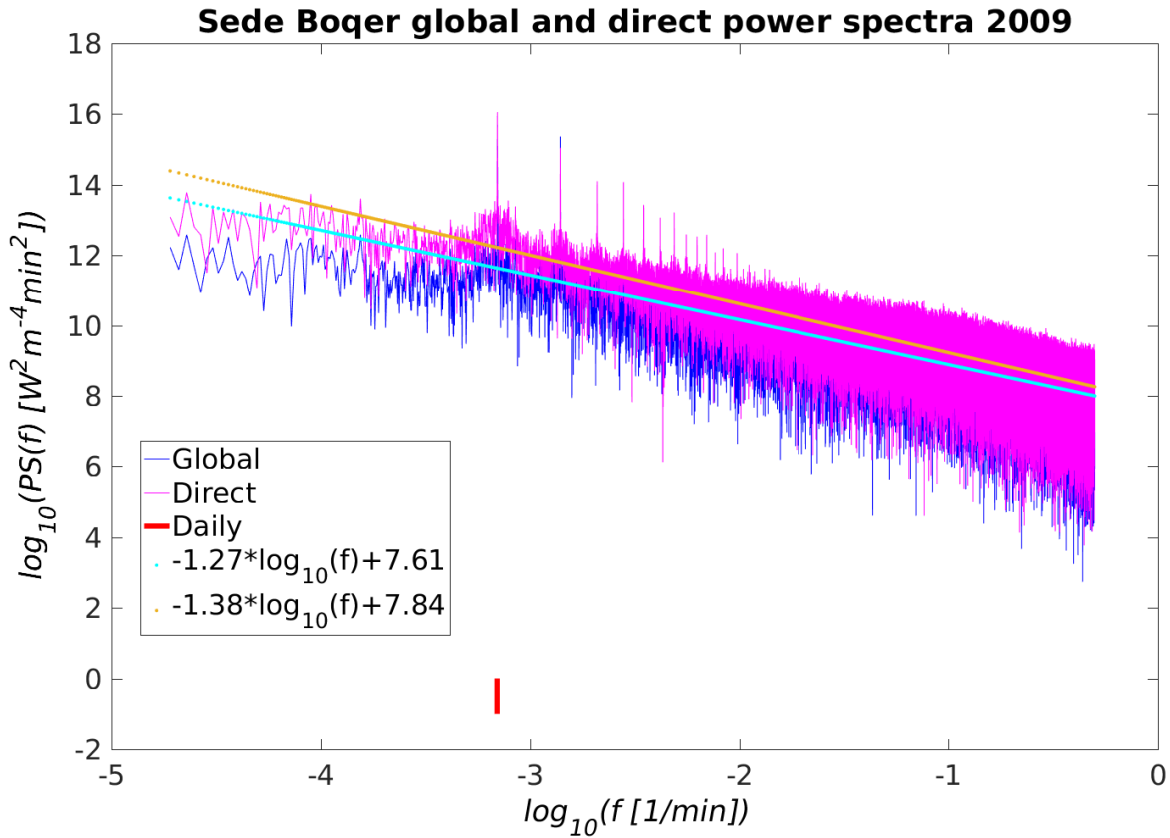}
  \caption{Measured spectra (solid line) and fits (dots) for global (blue) and direct (pink) radiation in Sede Boqer, Israel (year 2009); daily frequency is marked as a red vertical bar. Spectra of the two components have different power-law exponents ($\zeta$, slope of fit) but possess similar magnitudes (fit intercept). The difference between the power-law exponents ($\zeta$) is, however, statistically significant; the 95\% confidence interval for the global component exponent is [-1.28,-1.27] and [-1.39,-1.38] for the direct component $\zeta$.}
\label{fig:MBG_Direct_Measured_2009}
\end{figure}

\begin{figure}[!ht]
  \centering
  \includegraphics[width=\columnwidth]{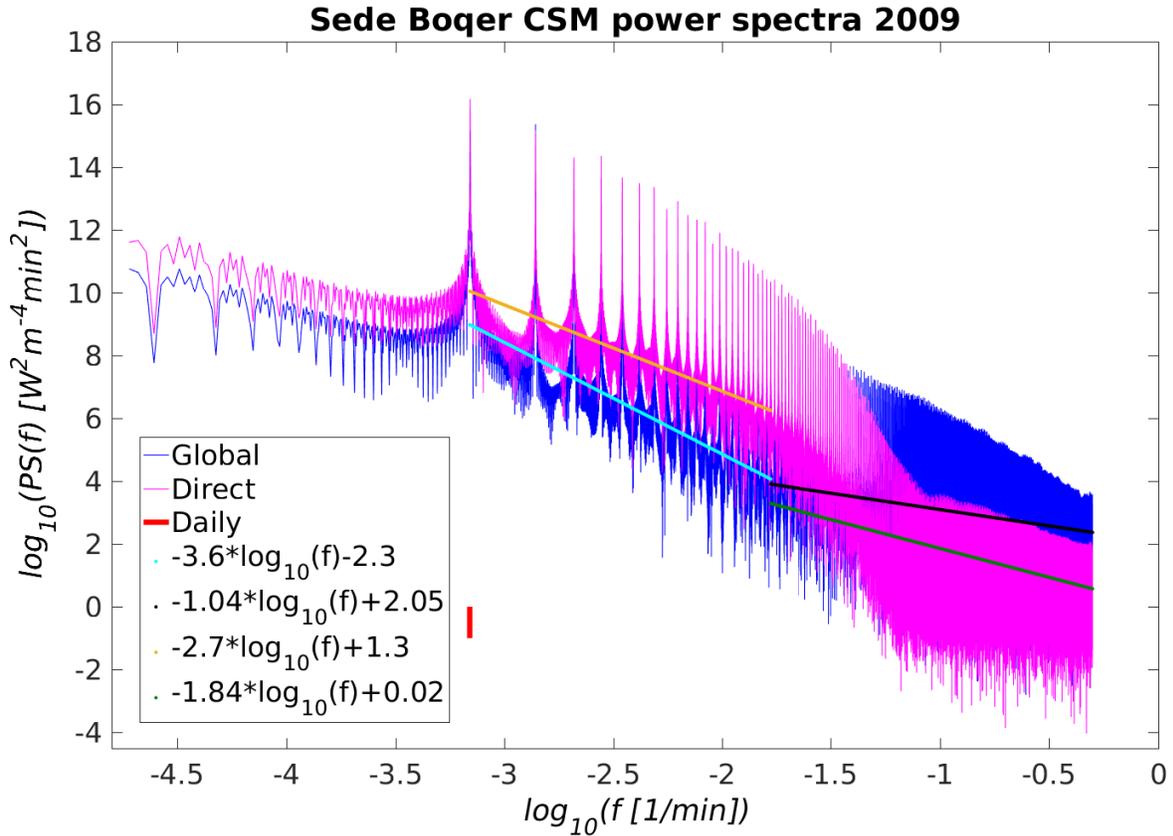}
  \caption{Calculated (CSM) spectra (solid line ) and fits (dots) for global (blue) and direct (pink) radiation for Sede Boqer in the year 2009. The frequency range was divided into high and intermediate frequency ranges; daily frequency is marked as a red vertical bar.}
  \label{fig:MBG-Ineichen-2009}
\end{figure}

As an example, Fig.~\ref{fig:MBG_Direct_Measured_2009} presents spectra for one year's data (2009) from sensors at Sede Boqer, Israel; also see Fig. S1 in the Supplemental Information (SI) for measurements during 2012. In addition to the expected peaks at $f = 1/$day and $f > 1/$day (due to radiation cutoff during dark hours) frequencies, we observe a continuous spectrum with a reasonably fit power-law decay for both direct and global components at $f > 1/$day; the fits do not vary much between 2009 and 2012. In contrast, the calculated (CSM) spectra for Sede Boqer in 2009 (Fig.~\ref{fig:MBG-Ineichen-2009}) are distinctly different (from the measured ones), particularly the deviation from a single power-law decay over the entire frequency range observed in both components (since they are not independent of each other). A reasonable fit is achieved only when done separately for the high ($f > 1/\rm{h}$) and intermediate (1/day $< f <$ 1/h) frequency ranges. This deviation can be understood from the fact that CSM assumes a clear sky and monthly varying turbidity as opposed to actual measurements that must contend with both clouds and faster turbidity variation.

\begin{figure*}[!ht]
  \centering
  \includegraphics[width=\linewidth]{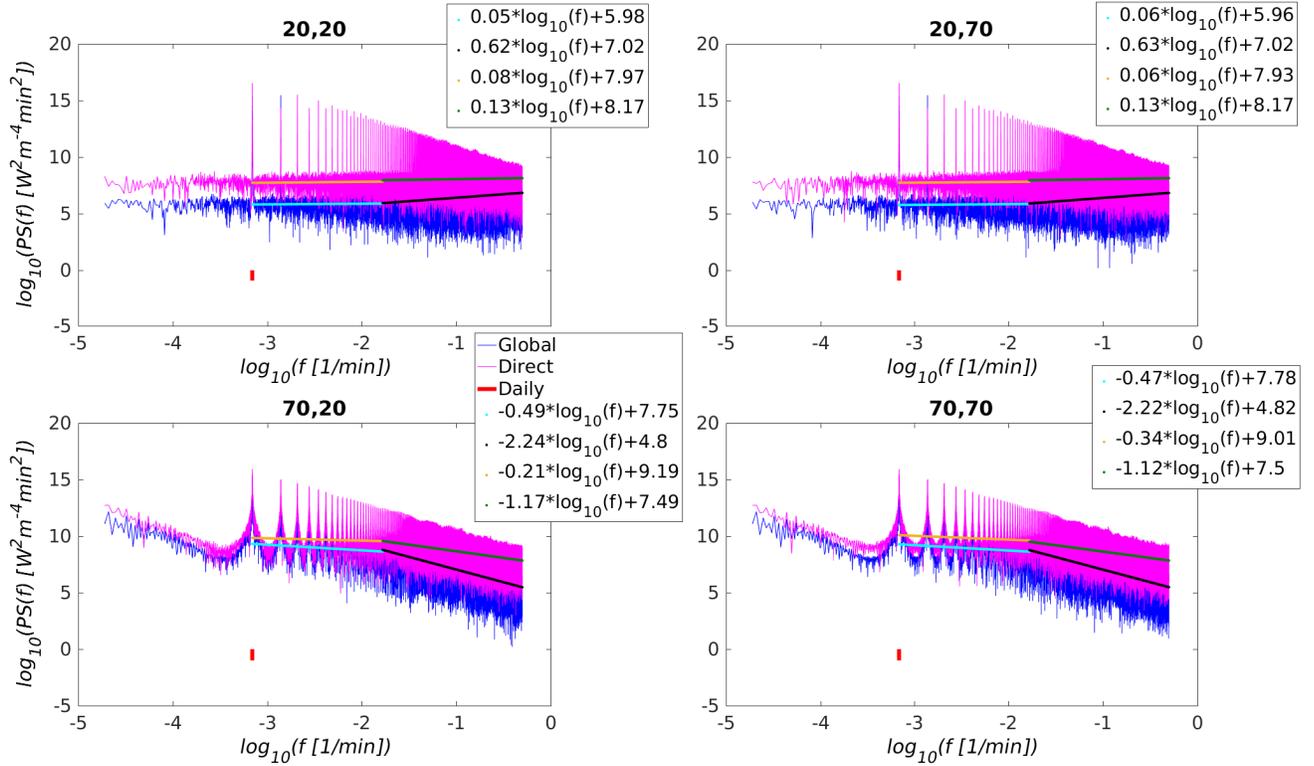}
  \caption{Calculated (CSMTL1) global (blue) and direct (pink) radiation spectra (solid lines) for the year 2015 with zero turbidity and their fits (dots) at four different locations with the following latitude and longitude coordinates: ($20^\circ,20^\circ$), ($20^\circ,70^\circ$), ($70^\circ,20^\circ$), and ($70^\circ,70^\circ$). The frequency range was divided into high and intermediate frequencies; daily frequency is marked as a red vertical bar.}
  \label{fig:2070comb}
\end{figure*}
\section{Results}
Having established the spectral power-law and its observable difference between measured and calculated spectra, we henceforth employ only calculated (CSM and CSMTL1) spectra in our analysis. This allows us to disentangle signal variability from local topography, cloud passage, etc. and focus exclusively on geographic dependence. To this end, we track how the exponent of the spectral power-law varies with geographic (latitude-longitude) location. To first demonstrate the dependence, in Fig.~ \ref{fig:2070comb}, we present CSMTL1 spectra, with elevation set to sea level to further simplify the analysis, at four different locations. Firstly, we observe strong variation in the spectral power-law exponent with location in both (high and intermediate) frequency regimes, and secondly, this variation exhibits a stronger dependence on the latitude than the longitude.

\begin{figure}[!ht]
  \centering
  \includegraphics[width=\columnwidth]{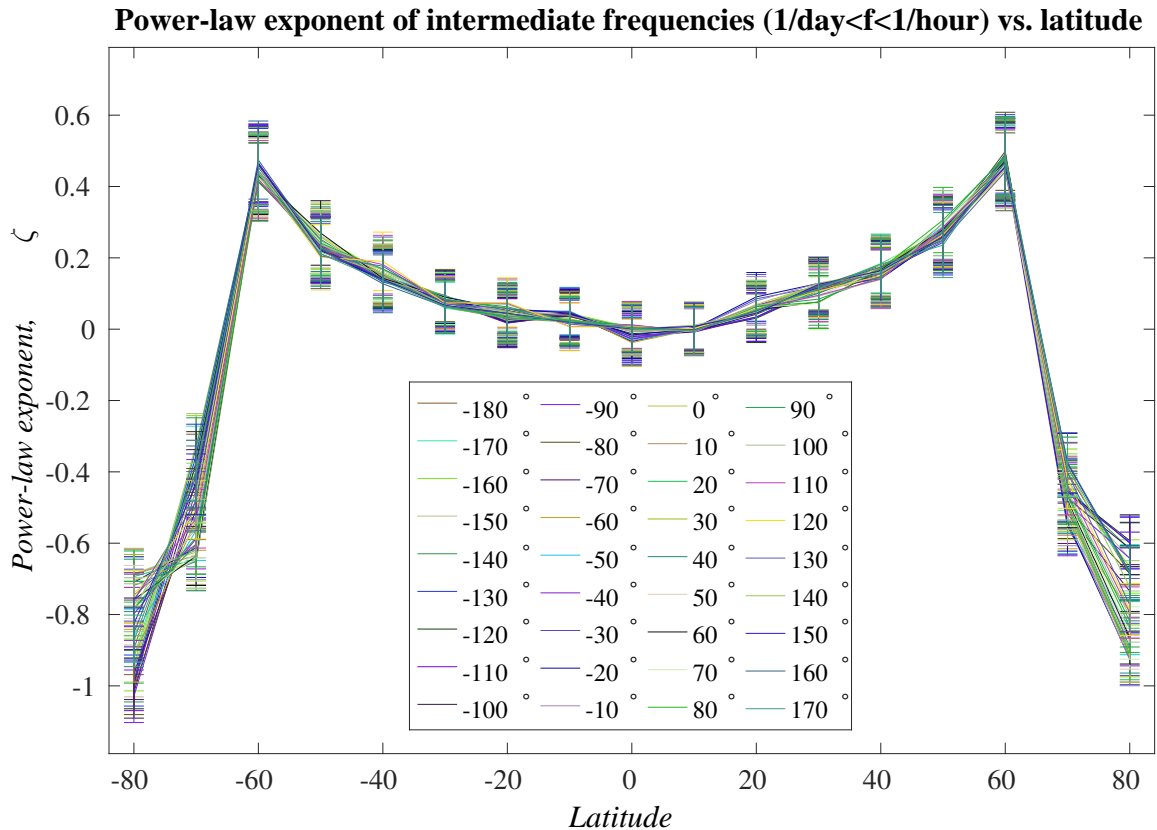}
  \caption{The latitudinal dependence of the power-law exponent of the intermediate frequency global radiation spectra. The power-law exponent is plotted against the latitude for each longitude. The latitude and the longitude vary in steps of $10^\circ$ in the range of latitudes $-80^\circ$-$80^\circ$ and longitudes $-180^\circ$-$170^\circ$. The different colors correspond to different longitudes as indicated. The error bars represent the 95\% confidence interval of the power-law exponent.}
  \label{fig:LFGlobalTL1_lat}
\end{figure}

To better understand this geographic dependence, we calculated the spectra at different coordinates for a range of latitudes $[-80^{\circ}, +80^{\circ}]$ and longitudes $[-180^{\circ} ,170^\circ]$ at $10^{\circ}$ intervals in each coordinate. Figure \ref{fig:LFGlobalTL1_lat} shows the power-law exponent, $\zeta$, from fit to the global radiation spectrum for intermediate frequencies at different latitudinal coordinates where each curve represents a different longitude. The error bars represent the 95\% confidence interval for the value derived from a linear fit to the logarithm of the power spectrum to the logarithm of the frequencies. It is apparent that the differences between the values of different longitudes are insignificant away from the poles as is the uncertainty interval. Closer to the poles, at latitudes $-80^\circ$ and $-70^\circ$ (above the Arctic Circle), and $70^\circ$ and $80^\circ$ (below the Antarctic Circle), the uncertainty is larger, yet the values for these latitudes are significantly different from those elsewhere; in these latitudes, the amplitudes of the low frequencies decay, while elsewhere, they either slowly grow or are almost constant. The observed non-monotonic dependence on latitude can be traced to the non-monotonic seasonal variations and daylight duration. The same information is presented in Fig.~S2 (see SI) but versus the longitude where each curve represents a different latitude. As expected, here too, no significant longitudinal dependence is observed in the power-law exponent except near the poles; the polar longitudinal dependence is, in fact, an artifact of the power-law fit at the low frequency end of the spectrum due to its non-contiguous nature as shown in the lower panels of Fig.~\ref{fig:2070comb}.

\begin{figure}[!ht]
  \centering
  \includegraphics[width=\columnwidth]{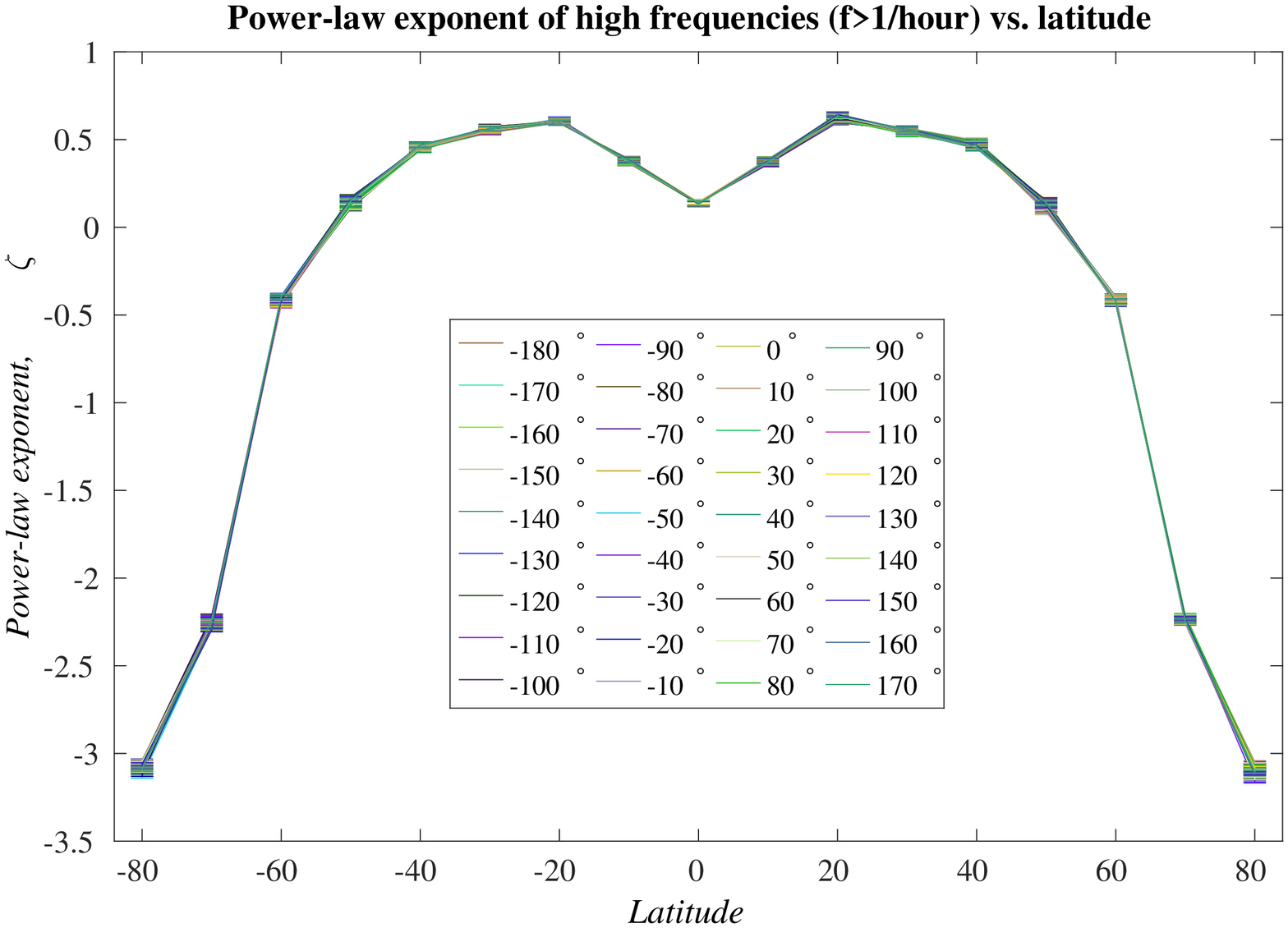}
  \caption{Spectral power-law exponent at high frequencies for global radiation; other details are the same as in Fig.~\ref{fig:LFGlobalTL1_lat}.}
  \label{fig:HFGlobalTL1_lat}
\end{figure}

Figure~\ref{fig:HFGlobalTL1_lat} (also see Fig.~S3 in SI) presents the variation in the power-law exponent versus the latitude at different longitudes for the high frequency end of the spectra. Once again, except near the poles, the high frequencies do not decay and, in fact, even grow with frequency (positive power-law exponent, $\zeta> 0$). Near the poles, the spectral decay is sharp on account of the weak daily cycle in these regions. The high frequency variation has no longitudinal dependence in any region, and the spread in individual curves is indeed less than the fit confidence intervals. Figures~S4--S7 present the direct component analog of Figs.~\ref{fig:LFGlobalTL1_lat}--\ref{fig:HFGlobalTL1_lat} (and Figs. S2--S3 in the SI) and exhibit more polar variation in longitudinal dependence at low frequencies than the global component. Away from the poles, the variation is low relative to the global component and almost constant in frequency, i.e., the power-law exponent is close to zero.

Having established the geographic dependence of the spectral power-law exponent, we now investigate the effect of geographic smoothing via two different averages. This smoothing analysis assumes significance particularly in the context of solar photovoltaic power production, where one is interested in the extent to which fluctuations are smoothed when power from geographically distributed photovoltaic power plants is summed at the grid. In the case of wind power, the largest scales in atmospheric flows correlate distant wind plants over hundreds of kilometers. When power from these geographically distributed plants is summed at the grid level, their fluctuations smooth until they reach a spectral limit of $f^{-7/3}$ \cite{Bandi2017}. Similarly, in solar photovoltaic power production, with the correlation length set by the sun, all power plants become globally correlated. Considering fluctuations from cloud passage alone, a theoretical limit for geographic smoothing in a photovoltaic power of $f^{-2}$ has been shown \cite{Klima2018}, but this limit does not take into account the fact that the clear sky spectrum (without clouds) already possesses a power-law dependence, which we exclusively consider here.

In the first method, an average was conducted using CSMTL1 data for the year 2015 over four locations in an arbitrarily selected latitudinal range ($29^\circ$--$32^\circ$) and $34^\circ$ longitude where spectral decay is minimal at low frequencies (Fig.~\ref{fig:LFGlobalTL1_lat}) and, in fact, increases at high frequencies (Fig.~\ref{fig:HFGlobalTL1_lat}) for both radiation components (see Fig.~S8 in the SI for the power spectra at the abovementioned locations). On averaging the radiation at these four locations, the spectral power-law exponent decreases only marginally for the global component in both frequency ranges, and for the direct component, the marginal decrease is observed only at high frequencies. Overall, averaging over four locations at different latitudes (all close to $30^\circ$), but at a fixed longitude, shows no significant spectral smoothing or modification for either radiation component. In the second averaging method, a similar scenario, at a fixed latitude ($30^\circ$) and five longitudinal locations ($32^\circ$--$36^\circ$) one degree apart, yielded similar results (see Fig. S9).
\begin{figure}[!ht]
  \centering
  \includegraphics[width=\columnwidth]{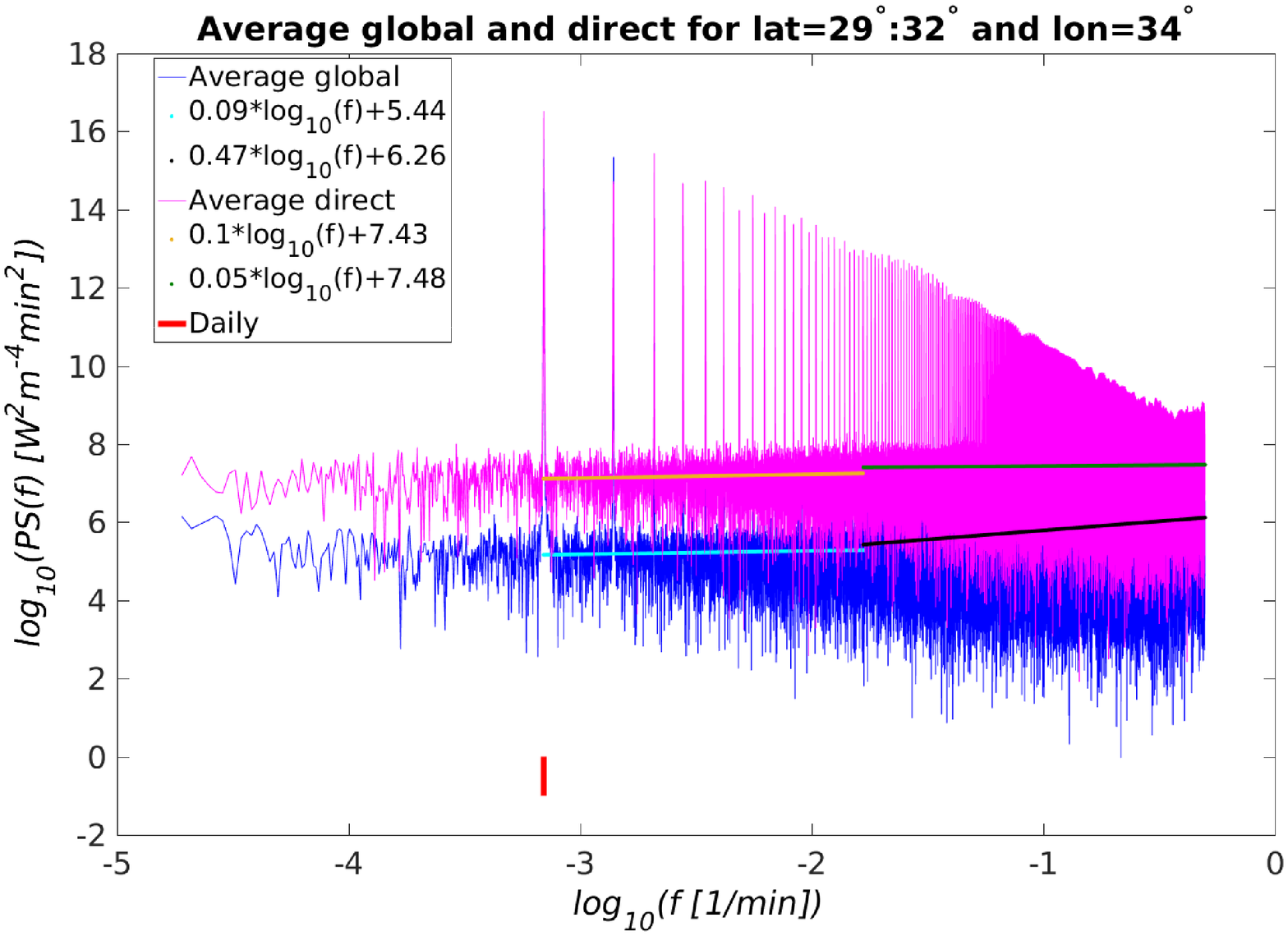}
  \caption{Global (blue) and direct (pink) radiation spectra (solid lines) for the averaged (over four locations in the latitudinal range $29^\circ--32^\circ$ and $34^\circ$ longitude) radiation with fits for intermediate and high frequency ranges.}
  \label{fig:psavglon34}
\end{figure}

\begin{figure}[!ht]
  \centering
  \includegraphics[width=\columnwidth]{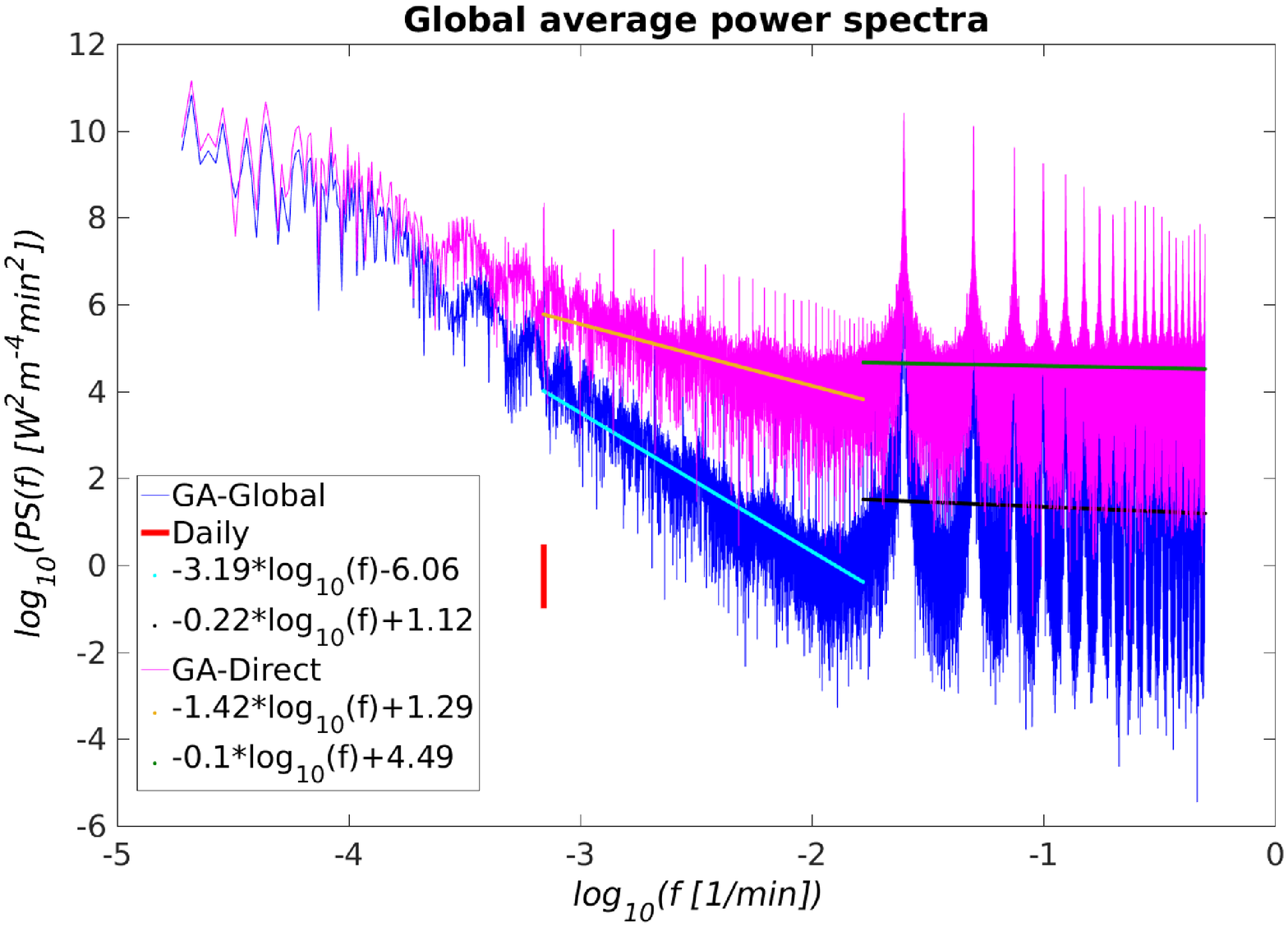}
  \caption{The power spectra of the global and direct components of the globally averaged radiation (see text for definition of the average).}
  \label{fig:IneichenGAGDps2015}
\end{figure}

Finally, we performed a global average of CSMTL1 data for the year 2015 over all global locations at which we calculated the radiation, which included 36 longitudinal values for each of 17 latitudinal values yielding a total of 612 sampling points. We note that this type of averaging corresponds to the average radiation of areas of equal size centered at the location of the grid points. The global and direct components spectra are presented in Fig.~\ref{fig:IneichenGAGDps2015}. The weakening of the daily frequency spectral peak is readily apparent owing to spatial averaging, and the difference in low and high frequency behavior observed at specific geographic locations is still captured in the global average. The spectral decay at intermediate frequencies does steepen for both components (with 95\% confidence intervals of the power-law exponent being (-3.24, -3.14) for the global component and (-1.46, -1.38) for the direct component), whereas the high frequency range spectra remain flat (with 95\% confidence intervals of the power-law exponent being (-0.24, -0.20) and (-0.11, -0.09) for the global and direct components, respectively).

\section{Discussion}

\begin{figure}[!ht]
  \centering
  \includegraphics[width=\columnwidth]{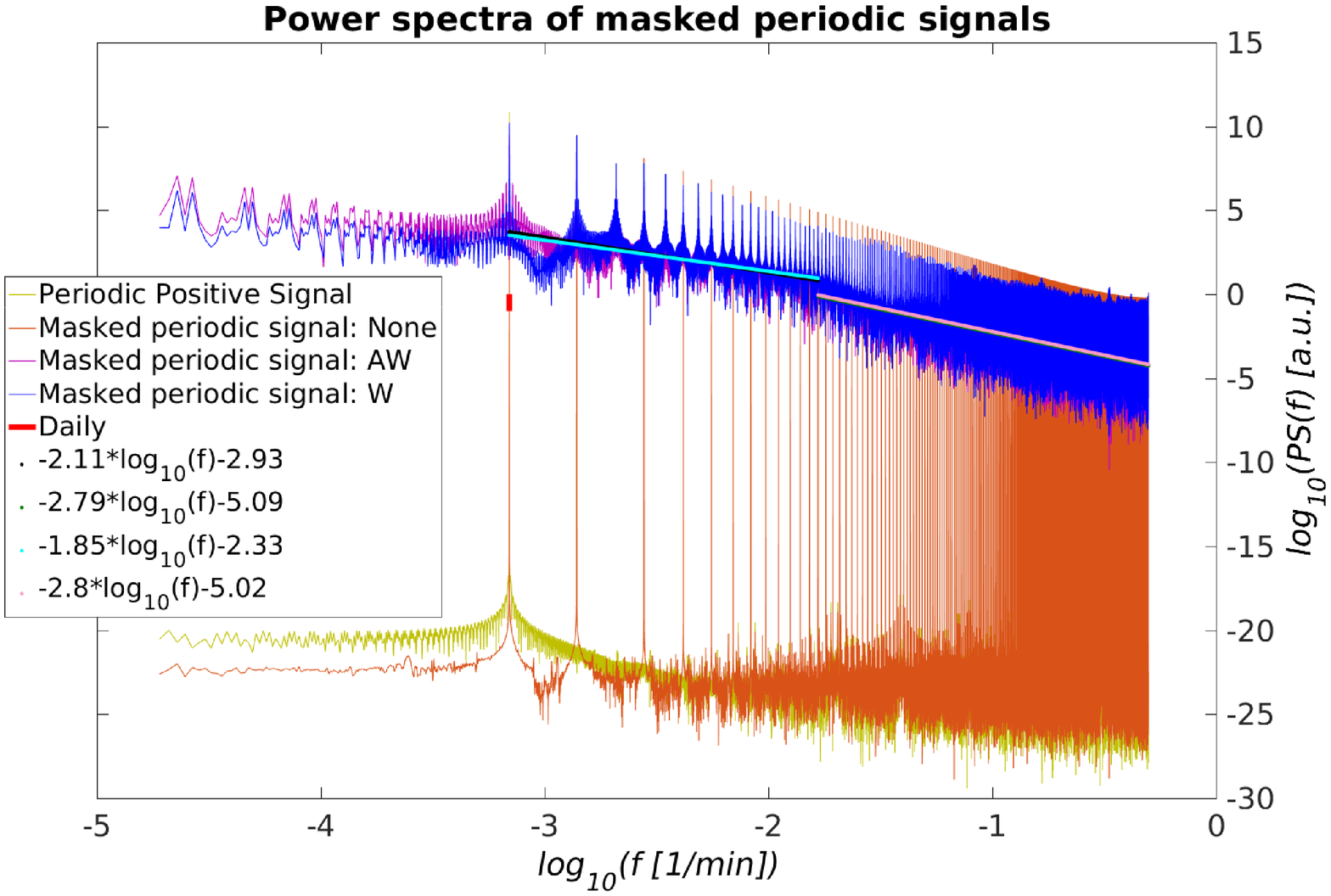}
  \caption{Power spectra for periodic signal with daily frequency (green-yellow line), masked periodic signal with fixed 12 hour on/off time (brown line), periodic signal with monthly varying “on time” according to daylight duration at $35^\circ N$ on the 16th of the corresponding calendar month and daily normalized amplitude equal to 1 (blue line), and masked periodic signal with monthly varying “on time” according to the daylight duration on 16th of each calendar month at $35^\circ N$ and also monthly varying amplitude as determined by the mask (see text for more details, purple line). The other lines present the power-law fits for the ``Masked periodic signal: W'' and ``Masked periodic signal: AW'' spectra at intermediate and high frequency ranges. }
  \label{fig:Mask}
\end{figure}

Having presented the results, we now explain the solar radiation spectrum starting from a simple temporally periodic signal, to which incremental additions of complex features reveal the minimal ingredients that explain the primary characteristics of the radiation spectra, {\it viz.} their power-law dependence, and their intermediate and high frequency behavior. First, we generated a yearlong time series with a 1 min time step of a periodic signal that is always positive, $cos(2\pi f_d t)+1$, where $f_d$ is the daily frequency (1/day), and, as expected, its spectrum (denoted as ``Periodic positive signal'' in Fig.~\ref{fig:Mask}) has a sharp peak at the daily frequency and some noise, which is essentially zero (note log scale of the figure). The second power spectrum shown in Fig. \ref{fig:Mask}, denoted as ``Masked periodic signal: None'', corresponds to the periodic signal $cos(2\pi f_d t)$ filtered such that negative values are set to zero, i.e., the signal is $max(cos(2\pi f_d t),0)$. This is the most basic attempt to describe a signal with a daily frequency and a ``dark'' period for half of each daily cycle. In this signal, both the amplitude and the width of the non-zero period remain constant. The spectrum of this simply masked signal is very similar to that of the ``Periodic positive signal'' with the main difference being the additional peaks at $f > 1/$day frequencies, which appear due to introduction of the ``dark'' periods.

Next we add an incremental feature to vary the non-zero (daylight) period and amplitude (for the soalr radiation, both quantities vary with season). We implemented it by using empirical data for daylight duration for different dates at $35^\circ$ latitude. For each month, we used the same daylight duration for the entire month and set it equal to the daylight time on the 16th of the corresponding calendar month. The positive periodic signal was modified by setting it to zero outside the daylight period and shifting it down such that the light starts from zero. This modification results in monthly variation of both the duration of the ``on time'' and of its amplitude (the maximal value during the ``on time''). This signal is denoted as ``Masked periodic signal: AW'' in Fig.~\ref{fig:Mask} and its spectrum readily varies from the ``Periodic positive'' and ``Masked periodic: None'' signals. In particular, both the power-law spectral dependence and the difference in low and high frequency behavior (showing different power-law exponents) come into sharp relief. To understand whether one needs both the variation of the ``on time'' and its amplitude in order to obtain the power-law dependence, we considered a fourth signal. This signal was similar to the third signal (``Masked periodic signal: AW''), but we normalized the signal such that in each day, the maximal value was equal to one. This fourth signal is denoted ``Masked periodic signal: W'' in Fig.~\ref{fig:Mask}. One sees that the power spectrum again exhibits a power-law dependence and two different exponents in the two regions of the frequency. Moreover, the power-law exponent for high frequencies is similar to the one for ``Masked periodic signal: AW'' but different at intermediate frequencies. This helps establish the fact that signals that vary more slowly than the daily frequency, {\it viz.} the daylight duration, control the spectral behavior in the intermediate and high frequency range.

These toy signals demonstrate that variation of daylight duration is sufficient to obtain a spectral power-law dependence, which in turn, explains its latitudinal variation. Armed with this understanding, it is worthwhile to explore its implications in various contexts. Since studies of climatic and ecological processes focus on the impact of solar radiation at the low (sub-daily) frequency end, spectra of relevant quantities in these fields at intermediate and high frequencies are lacking. Having established that the seasonal variation in daylight duration controls the observed behavior at intermediate frequencies, we posit the thesis that practitioners of climate and ecological sciences should pay attention to intermediate and high frequency variations and treat them on par with sub-daily frequencies.

Given that all renewable energy generation methods reflect the variability inherent in their energy sources \cite{Tabar2014, Bel2016, Bandi2017, Klima2015}, and since high to intermediate frequencies are of interest in renewable energy production and transmission, several studies have reported spectra in this frequency range for solar photovoltaic power production \cite{Ehnberg2005,Curtright2008,Lave2010,Marcos2011,Lave2012,Klima2015,Anvari2016}. For instance, the photovoltaic power spectrum at high frequencies $(0.001{\thinspace}{\rm Hz} < f < 0.05{\thinspace}{\rm Hz})$ (note that the daily frequency is $\approx10^{-5}\thinspace \rm{Hz}$ and the hourly frequency is $\approx3\times10^{-3}\thinspace \rm{Hz}$) was shown to follow the famous Kolmogorov \cite{K41} power-law decay with a -5/3 exponent at two different locations (Hawaii and Germany), and the -5/3 exponent was attributed to the turbulent advection of clouds above PV plants being reflected in the spectrum. Two sites in Arizona showed a $-2.6$ power-law decay of the solar power for the frequencies in the range of $\sim 10^{-6}-10^{-3}$ \rm{Hz} \cite{Curtright2008}. Another study using data from a power plant in Spain found that the PV plant acts as a low pass filter \cite{Marcos2011}. Namely, they showed that the measured irradiance spectrum follows a $-1.4$ power-law decay while the produced power spectrum follows the same power-law for low frequencies ($\sim 10^{-6}-10^{-3}$ \rm{Hz}) and a steeper decay with a power-law exponent of $-3.4$ for higher frequencies ($\sim 10^{-3}-10^{-1}\thinspace \rm{Hz}$). It was also suggested that the frequency at which the steeper decay starts may be a function of the plant size. An analysis of the power spectra of several PV power plants in India showed different power-law exponents in the range $-3.12$ to $-2.46$ and that connecting several plants in the same area together resulted in a steeper decay limited by the decay of the largest power plant \cite{Klima2015}.

Our analysis shows that the solar radiation spectrum already exhibits a robust power-law decay at intermediate and high frequencies even in the absence of extraneous factors such as turbidity and cloud occlusion. We therefore submit that while cloud occlusion certainly plays an important role in photovoltaic power fluctuations, this role should not be considered in a vacuum, but rather together with the bare solar radiation spectrum that a location experiences. This is particularly so because power fluctuations introduced by cloud passage act multiplicatively on the bare solar radiation. The spectra and their power-law exponents reported in the literature for photovoltaic power would be of little consequence unless compared against their counterparts for bare solar radiation at the same location.

\section{Conclusion}
In summary, we have studied the dependence of the solar radiation spectrum on different locations on the earth by employing the exponent of power-law dependence of the radiation spectrum as the quantity that captures the geographic dependence. Using a simple toy model comprising a periodic signal with a daily frequency, and by adding basic incremental features, we explain the origin of the power-law dependence of the radiation spectrum and its dependence on geographic location. Finally, we posit two theses: Firstly, experts in climate science and ecology should consider processes controlled by solar radiation at frequencies higher than the daily frequency because their character is determined by the variation of sub-daily frequencies. Secondly, research studies in photovoltaic power fluctuations should consider comparing their measured photovoltaic power spectra against bare solar radiation spectra to better understand the role of clouds and how they change the spectral character.

\section*{Acknowledgments}
MMB was supported by the Nonlinear and Non-equilibrium Physics Unit, OIST Graduate University. MMB gratefully acknowledges generous hosting by Prof. S. Sengupta at TCIS, TIFR Hyderabad while preparing this manuscript.

\bibliography{SummaryBib2}

\section{Supplementary Information}

\renewcommand{\thefigure}{{S\arabic{figure}}}
\subsection{Global radiation component spectrum}
\setcounter{figure}{0}

\begin{figure}[!ht]
  \centering
  \includegraphics[width=\columnwidth]{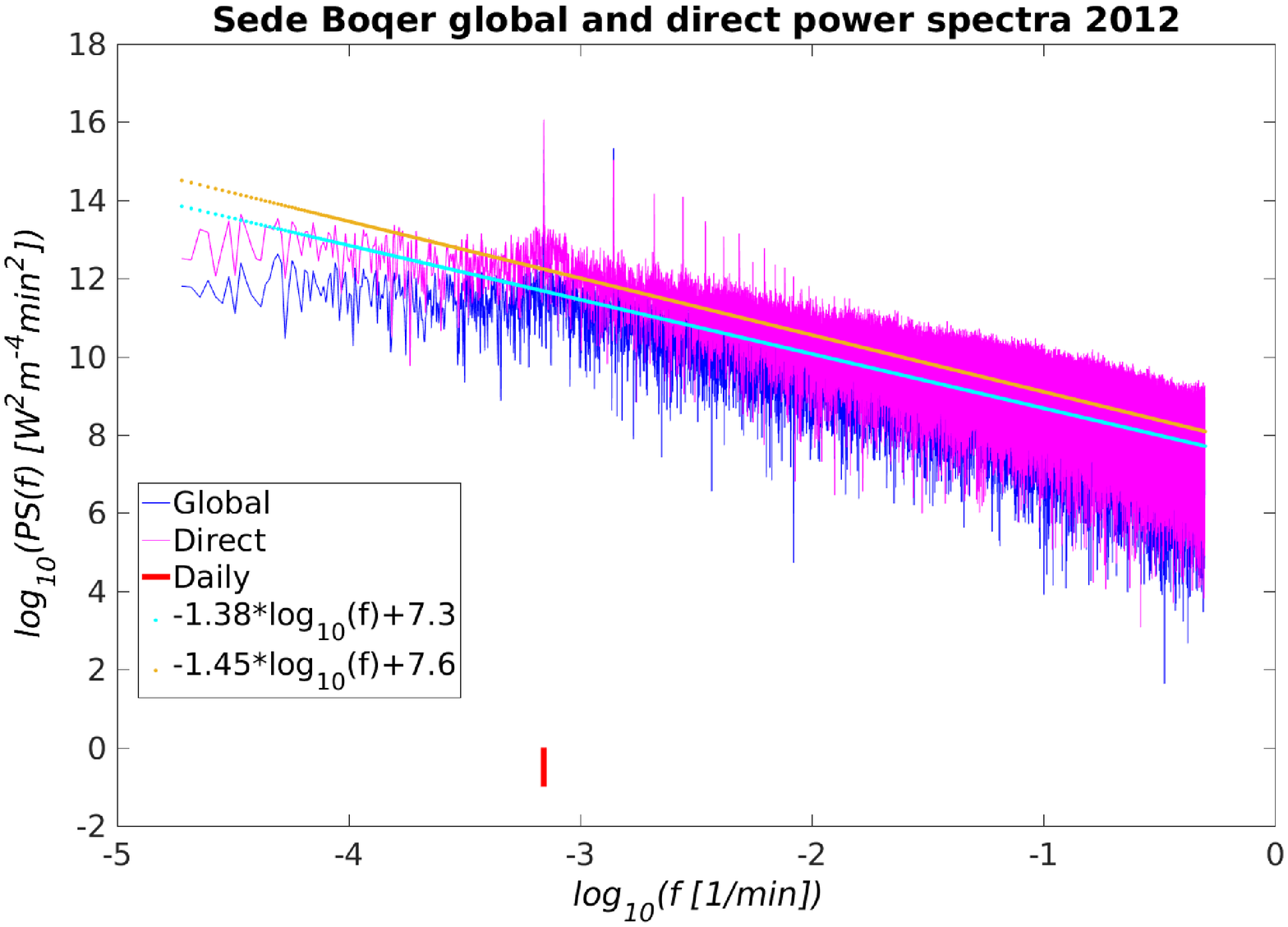}
  \caption{Spectra for the global and direct radiation components in Sede Boqer during the year 2012. The information and notation is identical to Fig. 1 of the main text. The 95\% confidence interval for the decay power is [-1.39,-1.38] for the global and [-1.46,-1.45] for the direct component.}
  \label{fig:MBG_Direct_Measured_2012}
\end{figure}

Figure \ref{fig:MBG_Direct_Measured_2012} presents the power spectra of the global and direct components of the solar radiation measured at Sede Boqer, Israel (see the main text for more details on the location) in the year 2012. The power-law fit is similar to the one found for the radiation measured in 2009 (presented in Fig. 1 of main text), as expected.

\begin{figure}[!ht]
  \centering
  \includegraphics[width=\columnwidth]{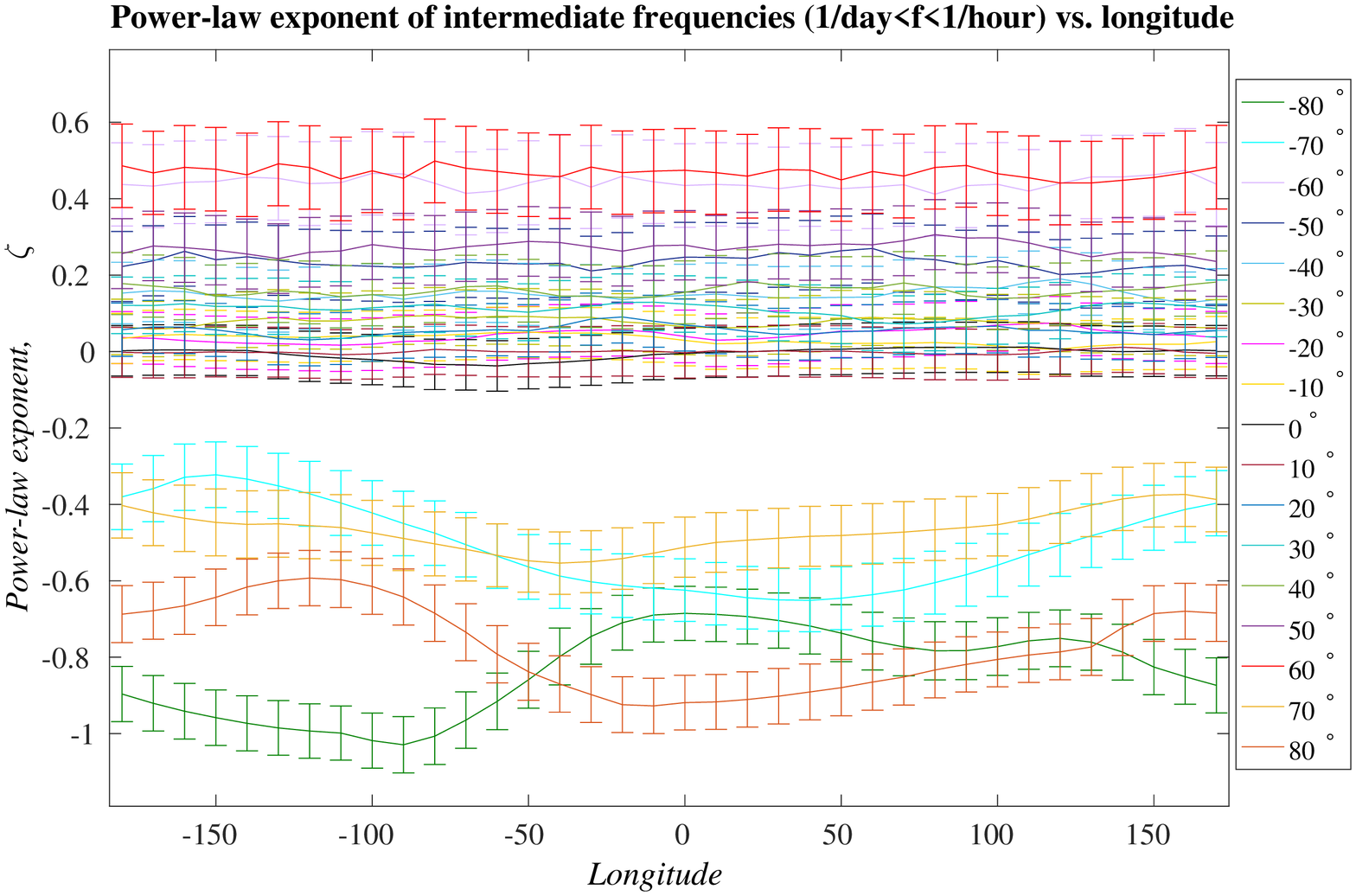}
  \caption{The power-law exponent of the intermediate frequency range global radiation spectra plotted against the longitude for each latitude. The other details are the same as in Fig. 4 of the main text.}
  \label{fig:LFGlobalTL1_lon}
\end{figure}

Figure \ref{fig:LFGlobalTL1_lon} presents the dependence of the power-law exponent of the intermediate frequency range on the longitude for different latitudes. This same information is also presented in Fig. 4 of the main text, but against the latitude instead of the longitude.

\begin{figure}[!ht]
  \centering
  \includegraphics[width=\columnwidth]{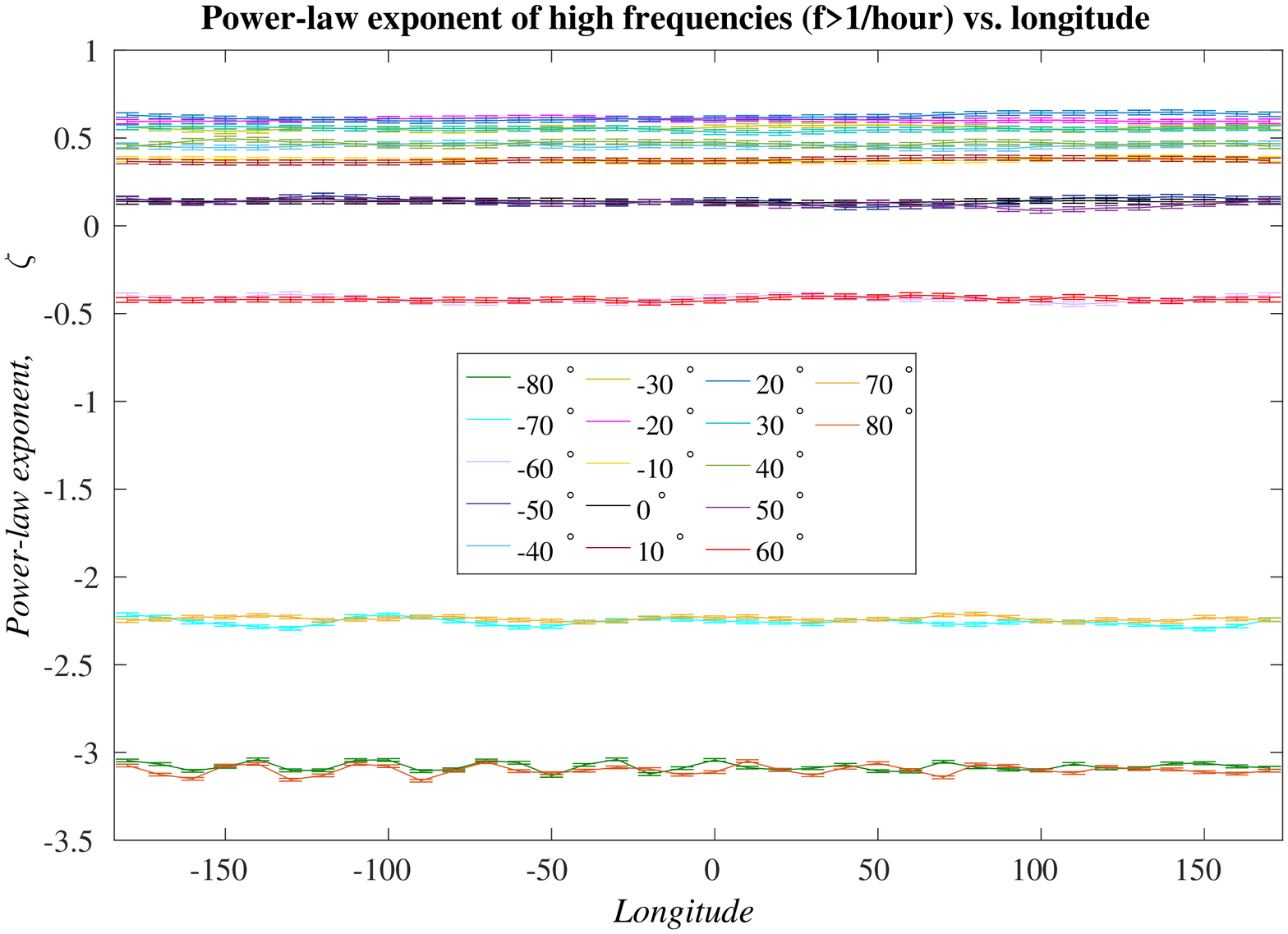}
  \caption{The dependence of the power-law exponent of the high frequency global radiation spectra on the longitude. Other details are the same as in Fig. \ref{fig:LFGlobalTL1_lon}.}
  \label{fig:HFGlobalTL1_lon}
\end{figure}

Figure \ref{fig:HFGlobalTL1_lon} presents the power-law exponent for the high frequency range versus the longitude. This figure presents the same information provided in Fig. 5 of the main text but versus the longitude instead of the latitude. 

\subsection{Direct radiation component spectrum}
\begin{figure}[!ht]
  \centering
  \includegraphics[width=\columnwidth]{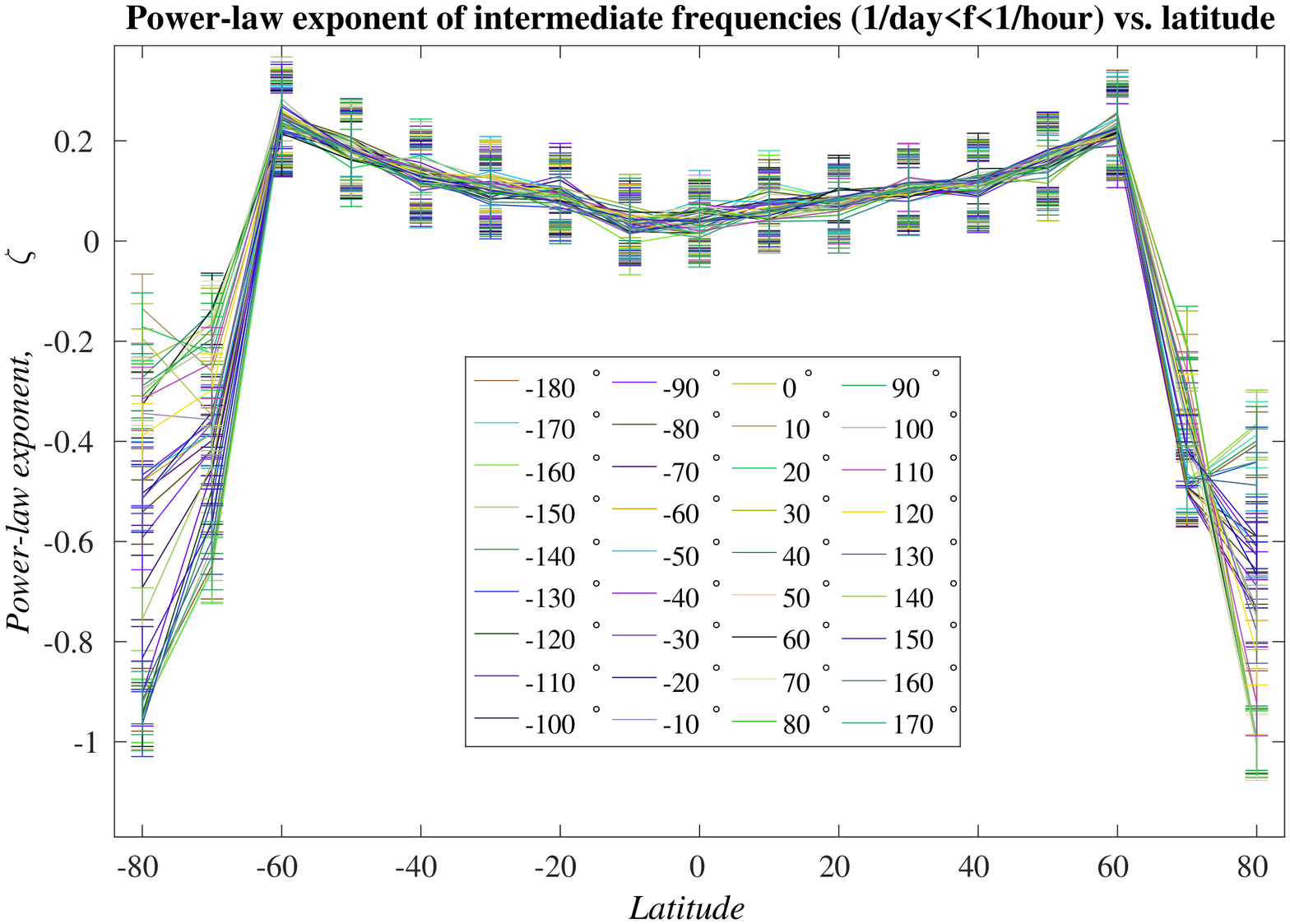}
  \caption{The power-law exponent at intermediate frequencies for the direct radiation spectra. The power-law exponent is plotted against the latitude for each longitude. The latitude and the longitude vary in steps of $10^\circ$ in the range of latitudes $-80^\circ$--$80^\circ$ and longitudes $-180^\circ$--$170^\circ$. The different colors correspond to different longitudes as indicated. The error bars represent the 95\% confidence interval of the power-law exponent.}
  \label{fig:LFDirectTL1_lat}
\end{figure}

Figure \ref{fig:LFDirectTL1_lat} shows that near the poles, the power-law exponent of intermediate frequencies for the direct radiation spectra varies more than it does for the global component. In the other latitudes, it varies less than it does for the global component, and it is almost constant with respect to the frequency (the power-law exponent is close to zero).

\begin{figure}[!ht]
  \centering
  \includegraphics[width=\columnwidth]{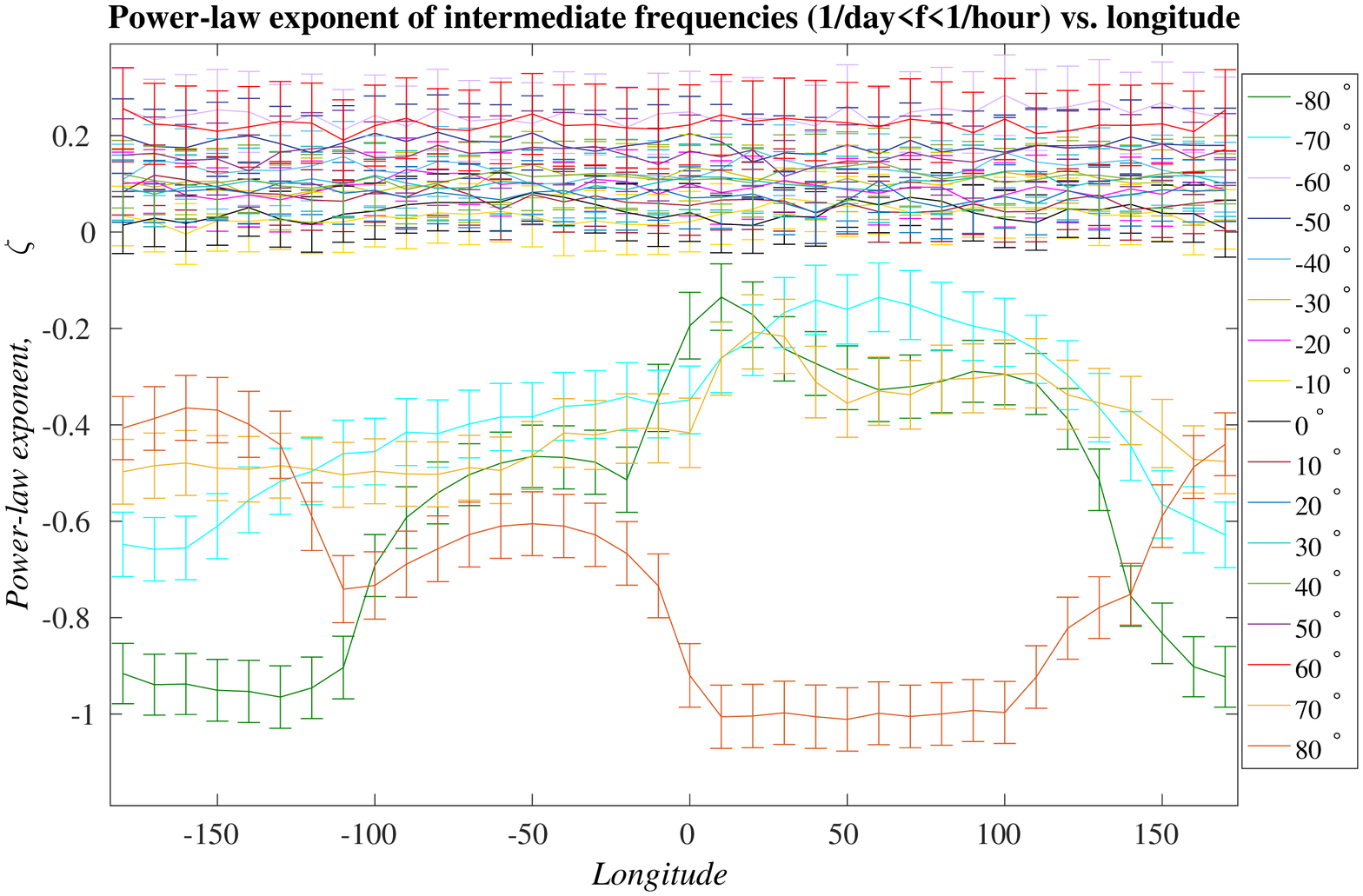}
  \caption{The power-law exponent at intermediate frequencies for the direct radiation spectra. The power-law exponent is plotted against the longitude for each latitude. Other details are the same as in Fig. \ref{fig:LFDirectTL1_lat}.}
  \label{fig:LFDirectTL1_lon}
\end{figure}

Figure \ref{fig:LFDirectTL1_lon} better illustrates the longitudinal dependence of the power-law exponent close to the poles and the lack of longitudinal dependence and almost flat spectra elsewhere.

\begin{figure}[!ht]
  \centering
  \includegraphics[width=\columnwidth]{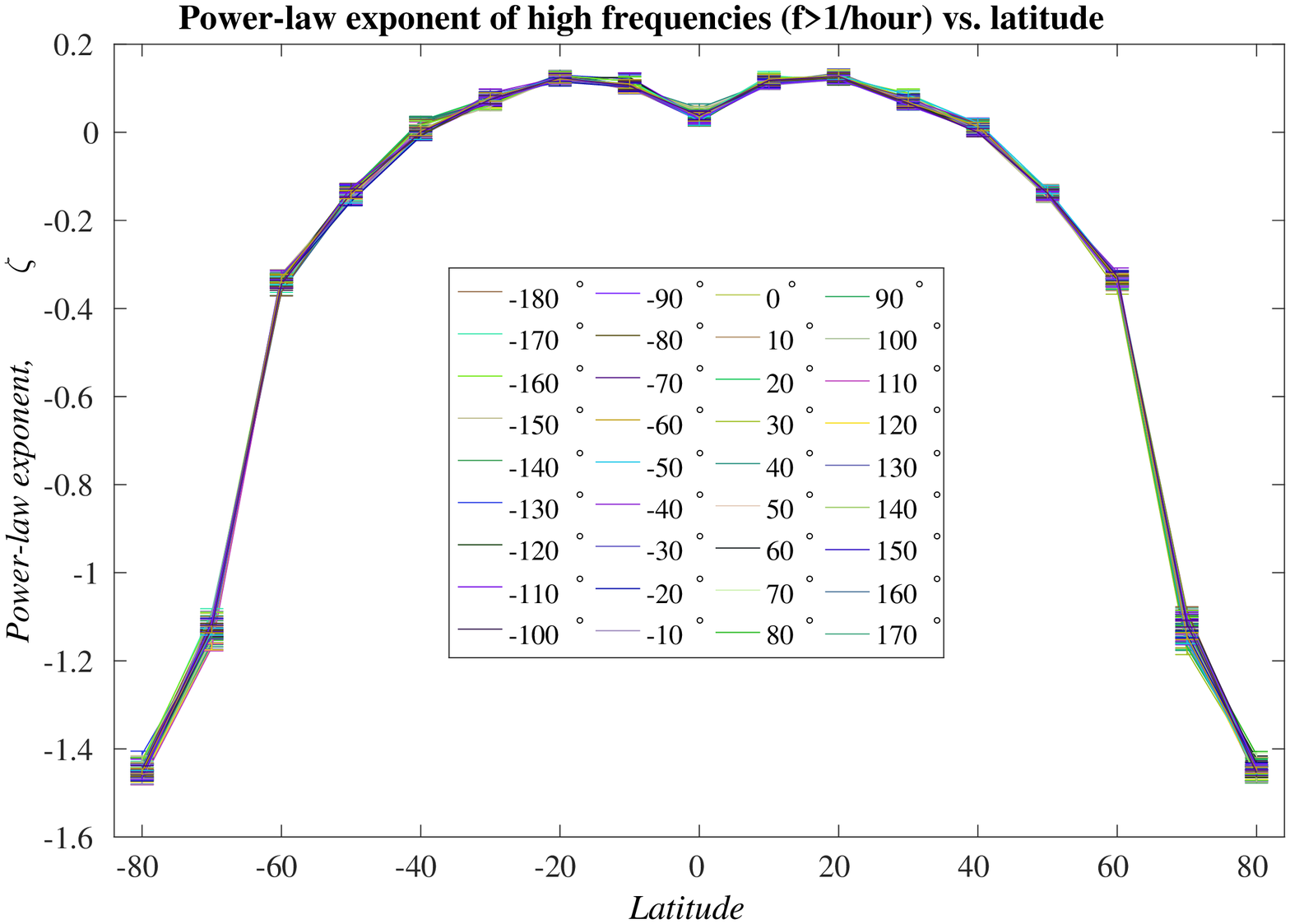}
  \caption{The dependence of the power-law exponent for high frequencies of the direct radiation spectra. Other details are the same as in Fig. \ref{fig:LFDirectTL1_lat}.}
  \label{fig:HFDirectTL1_lat}
\end{figure}

The power-law exponent at high frequencies for direct radiation spectra (Fig. \ref{fig:HFDirectTL1_lat}) shows a similar shape (as a function of the latitude) as the global component, but the values are different. 

\begin{figure}[!ht]
  \centering
  \includegraphics[width=\columnwidth]{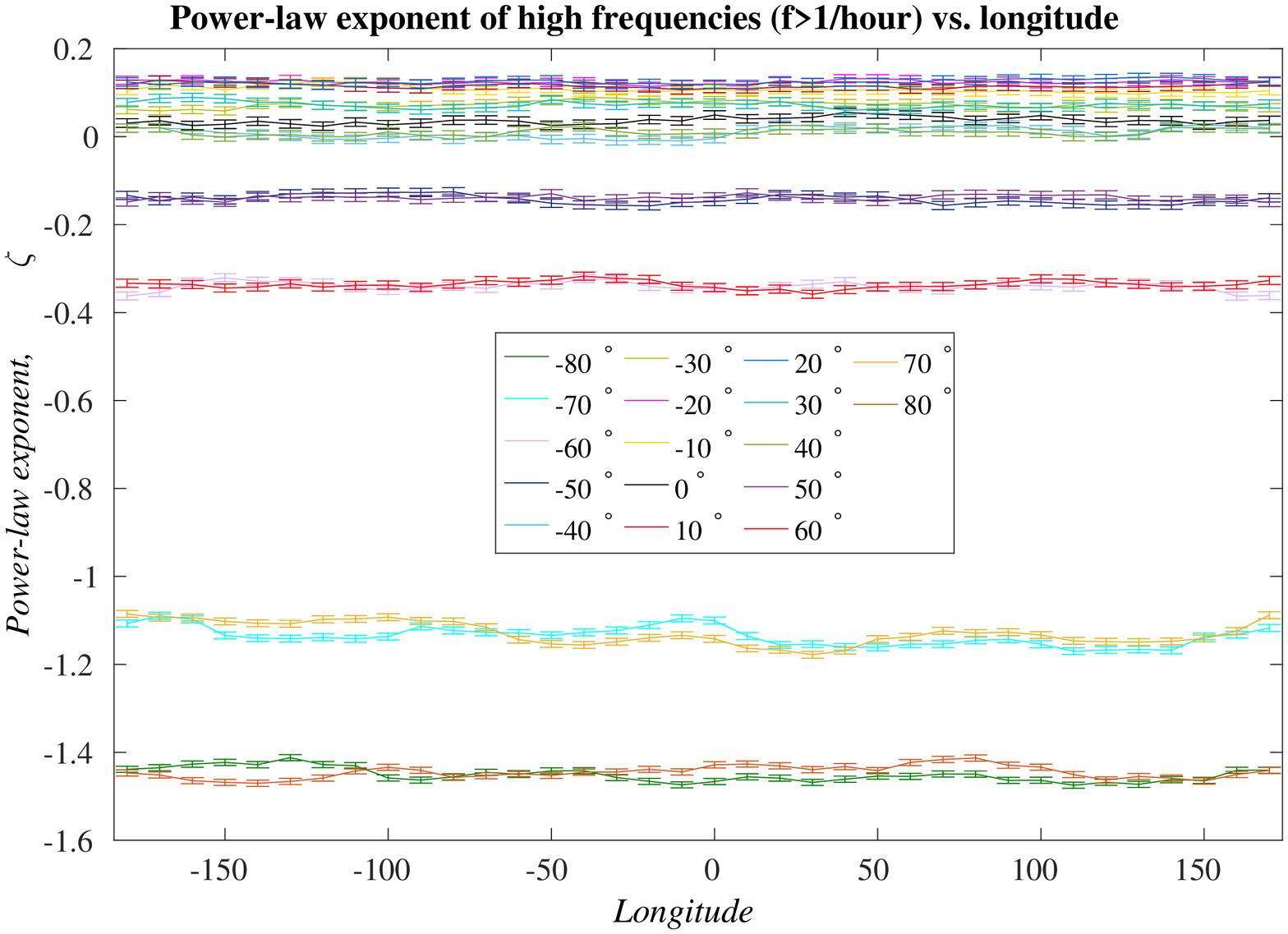}
  \caption{The longitudinal power-law exponent dependence of the high frequency direct radiation spectra. Other details are the same as in Fig. \ref{fig:LFDirectTL1_lon}.}
  \label{fig:HFDirectTL1_lon}
\end{figure}

Figure \ref{fig:HFDirectTL1_lon} better illustrates the lack of longitudinal dependence of the direct radiation spectral power-law exponent for high frequencies.

\subsection{Spatial averaging over different longitudes}
\begin{figure*}[!ht]
  \centering
  \includegraphics[width=0.45\columnwidth]{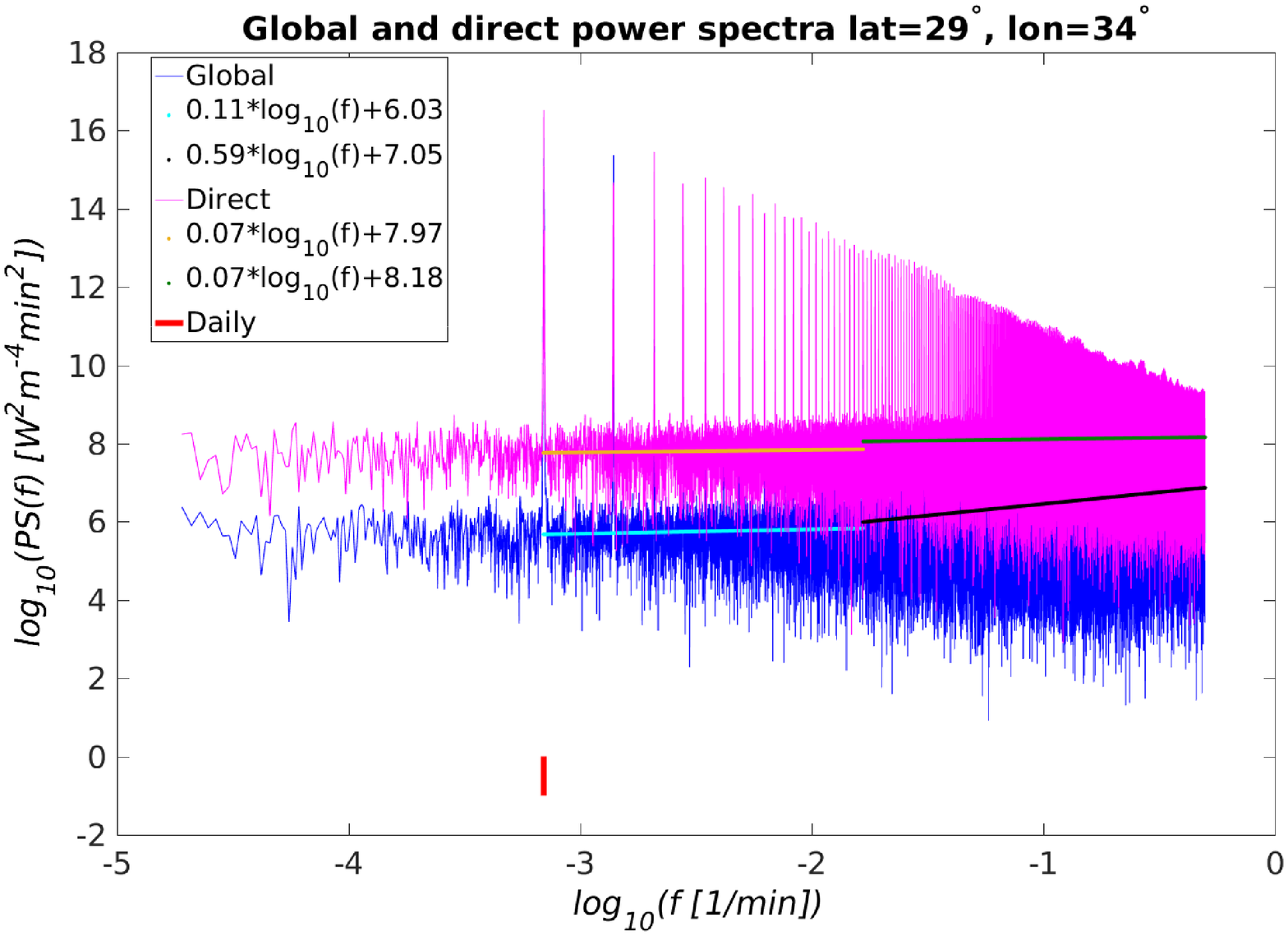}
  \includegraphics[width=0.45\columnwidth]{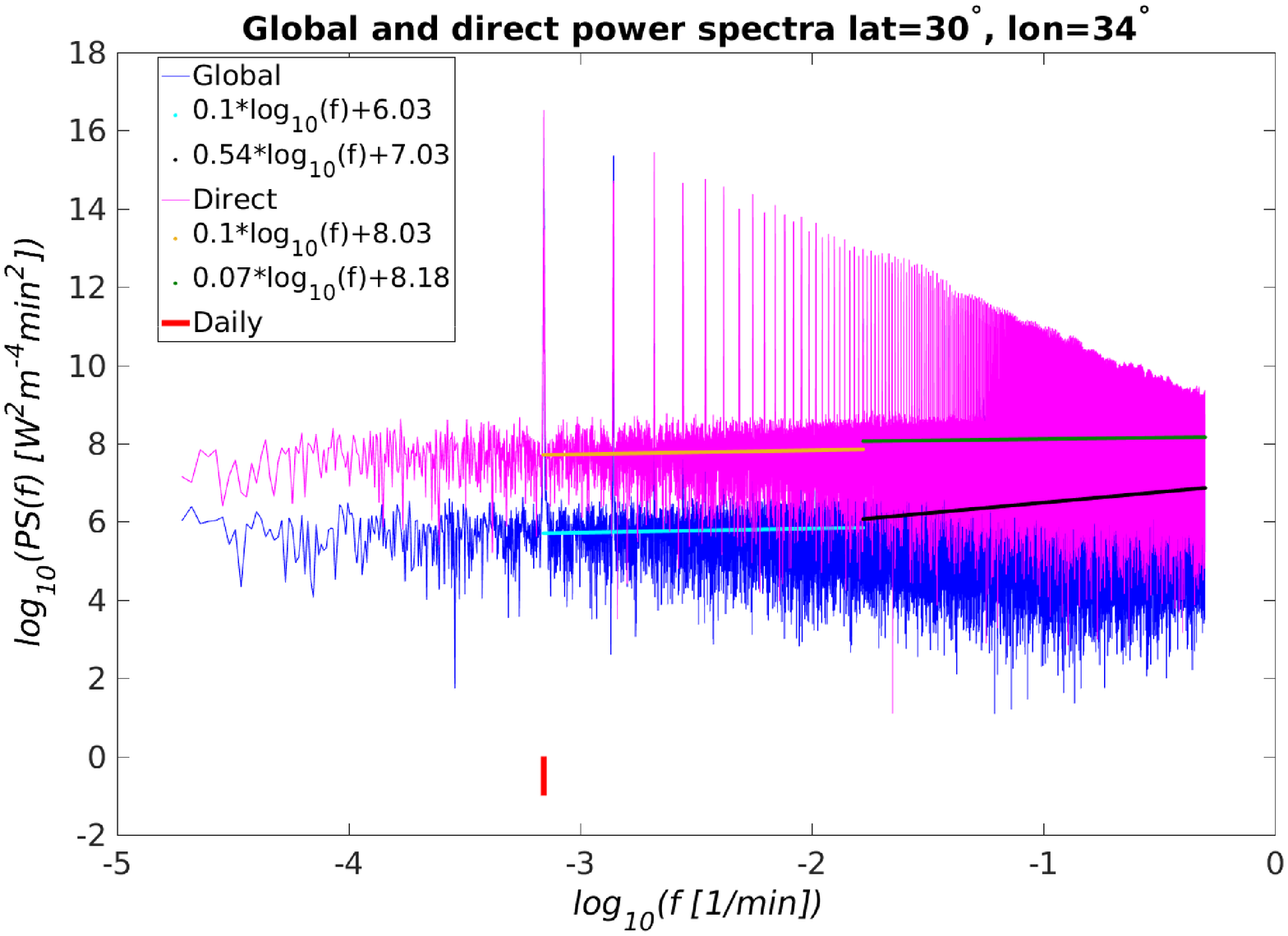}
  \includegraphics[width=0.45\columnwidth]{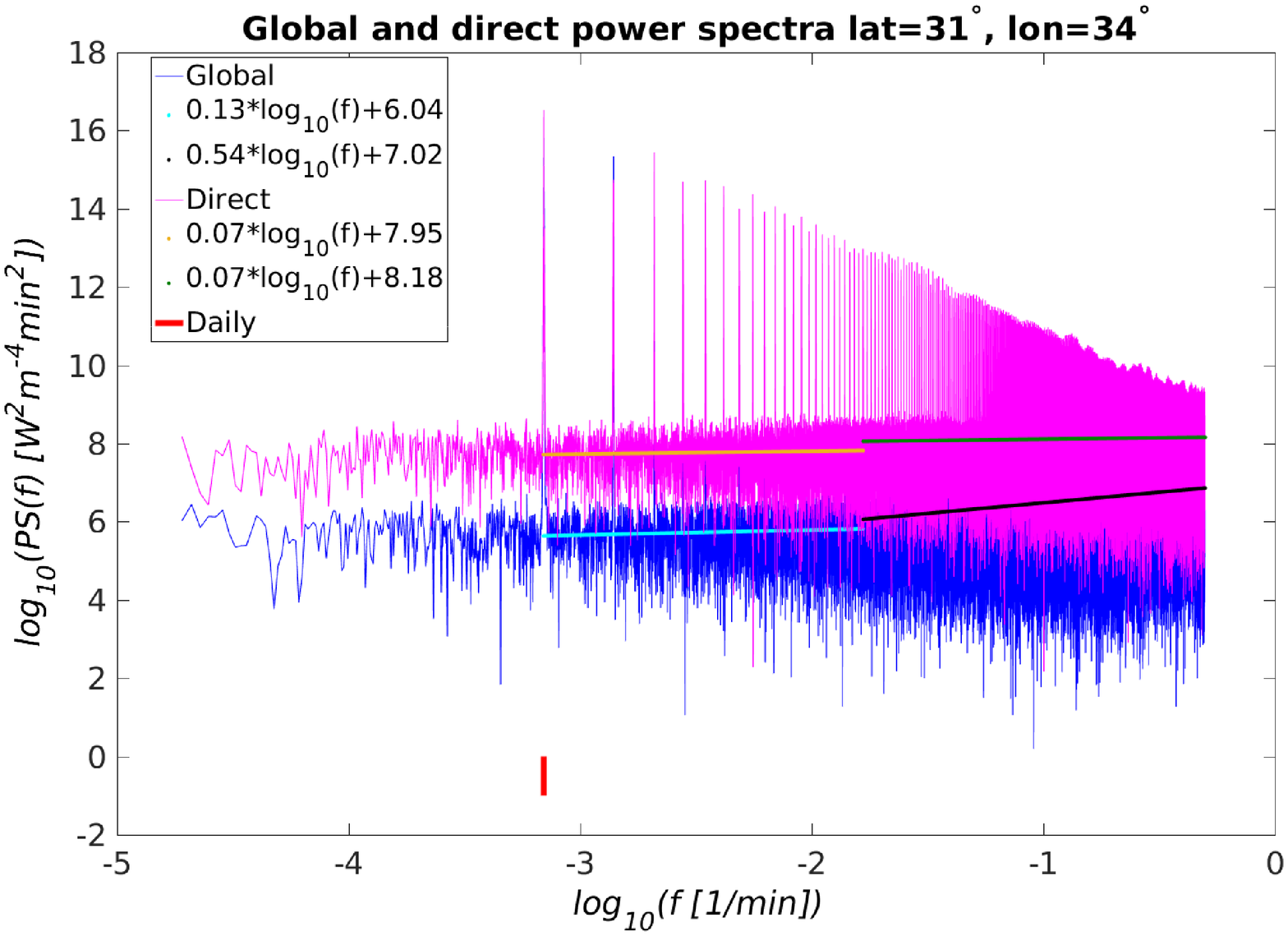}
  \includegraphics[width=0.45\columnwidth]{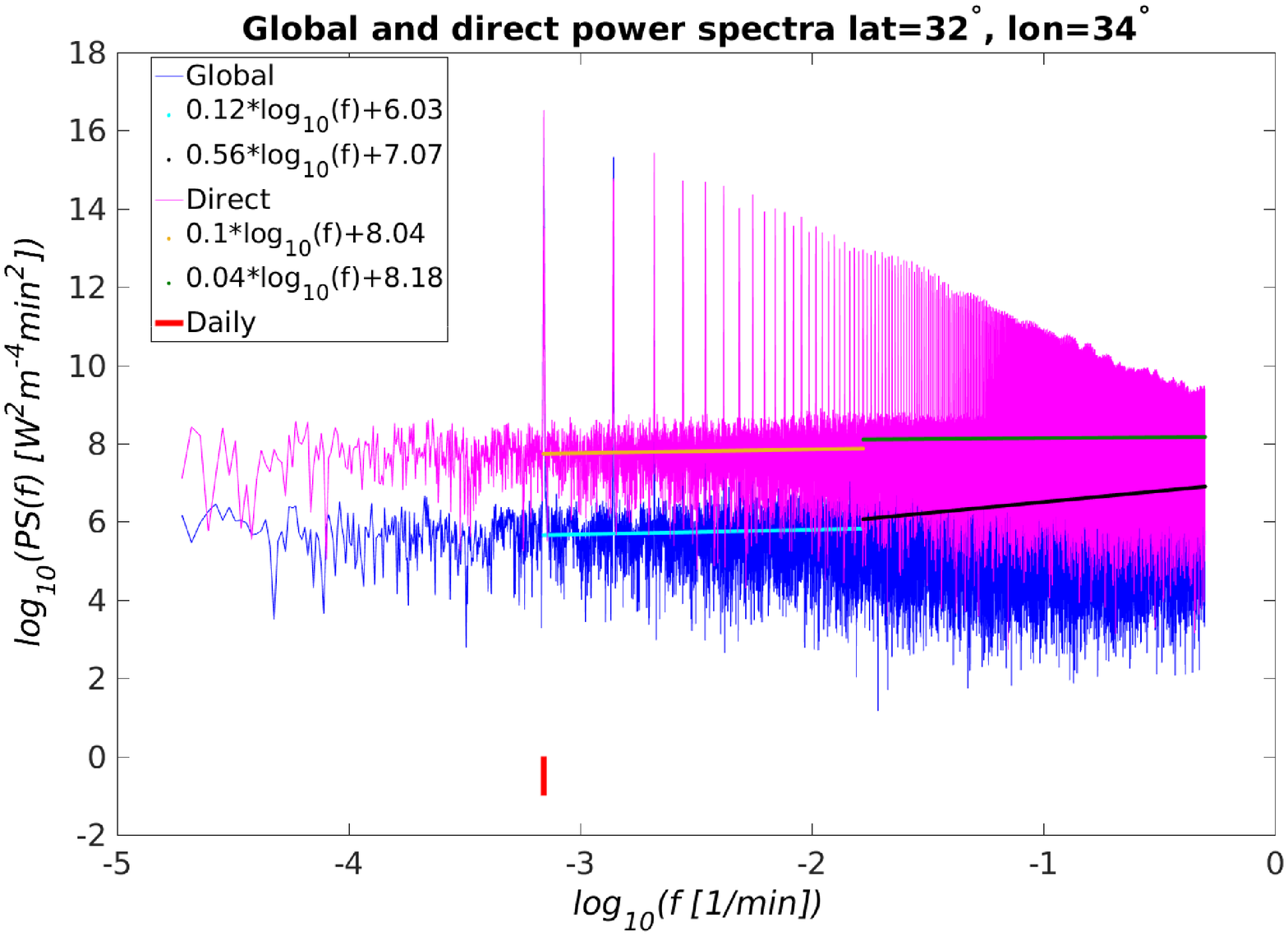}
  \caption{The power spectra of the global and direct components at four different locations with the same longitude, $34^\circ$, and latitudes varying by one degree from $29^\circ$--$32^\circ$.
  We present the power spectra of each component and the power-law fit at the two different frequency ranges.}
  \label{fig:pslon34}
\end{figure*}

Figure \ref{fig:pslon34} presents the power spectra for the four locations considered for the latitudinal averaging presented in Fig. 6 of the main text.

The power spectrum of the average over different locations with different longitudes, with a one degree interval between $32^\circ$ and $36^\circ$, but the same latitude, $30^\circ$. We found a small smoothing effect similar to what we found for an average over locations with different latitudes and the same longitude (see Fig. 6 of the main text).
\begin{figure*}[!ht]
  \centering
  \includegraphics[width=0.45\columnwidth]{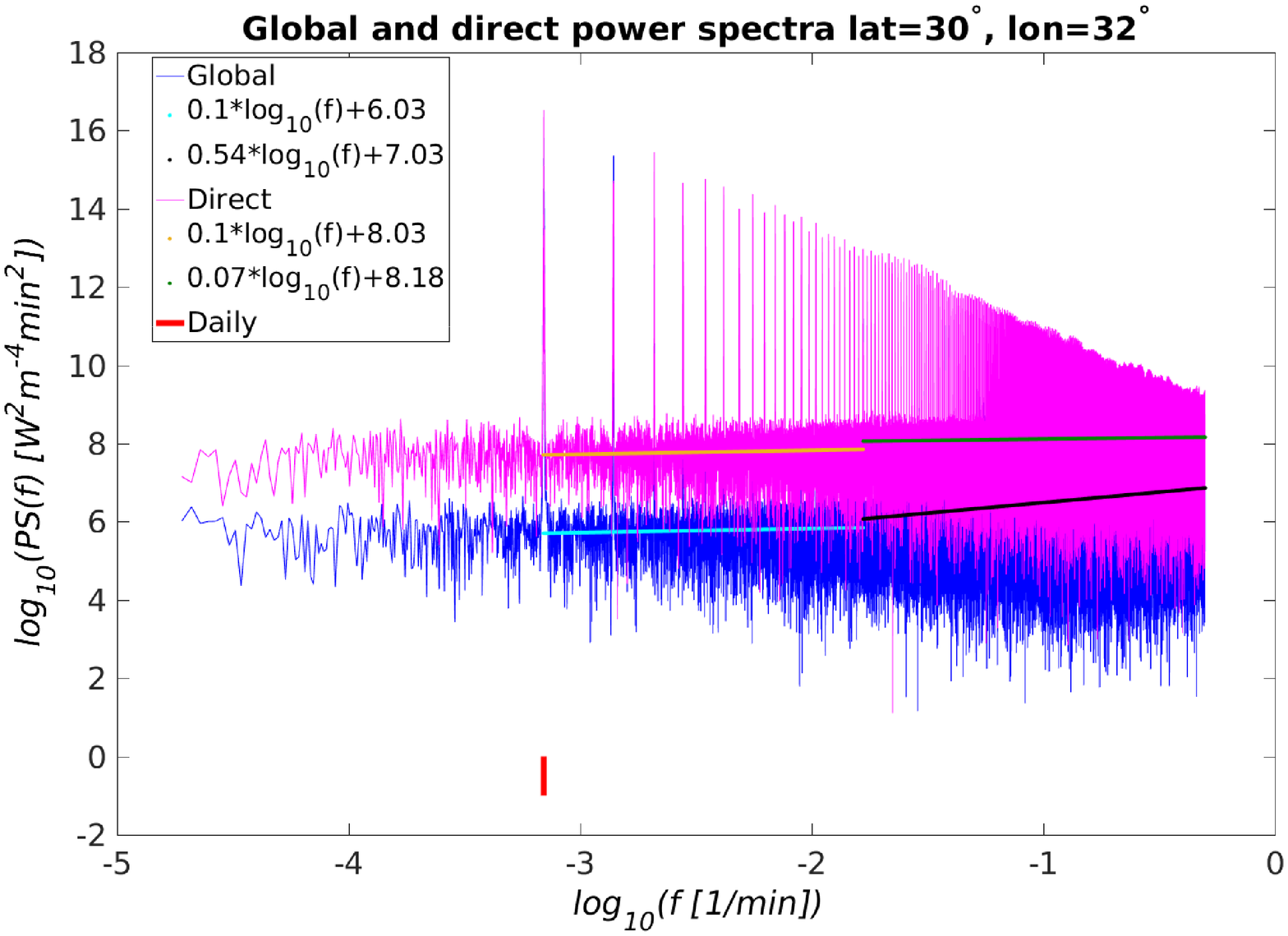}
  \includegraphics[width=0.45\columnwidth]{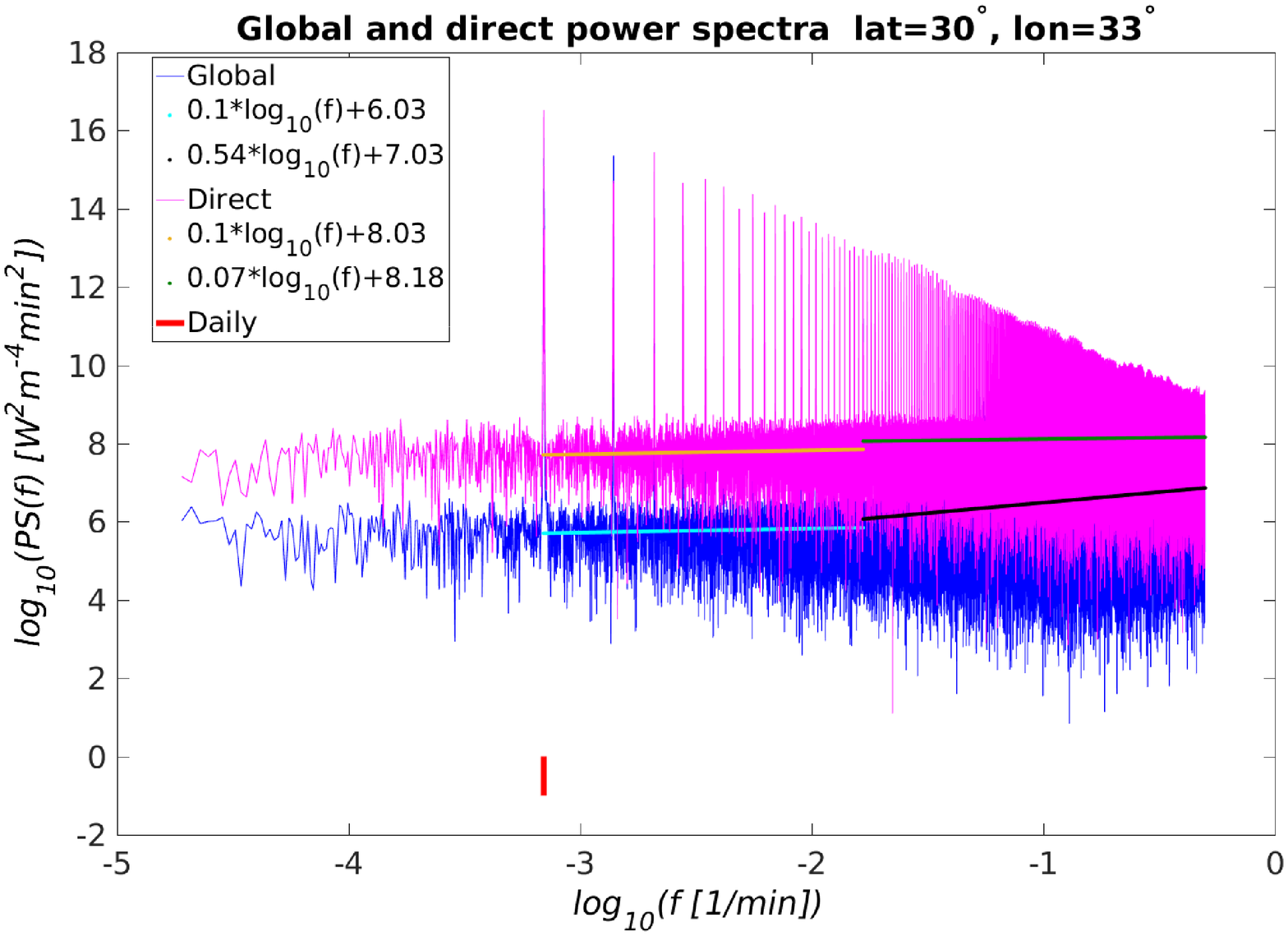}
  \includegraphics[width=0.45\columnwidth]{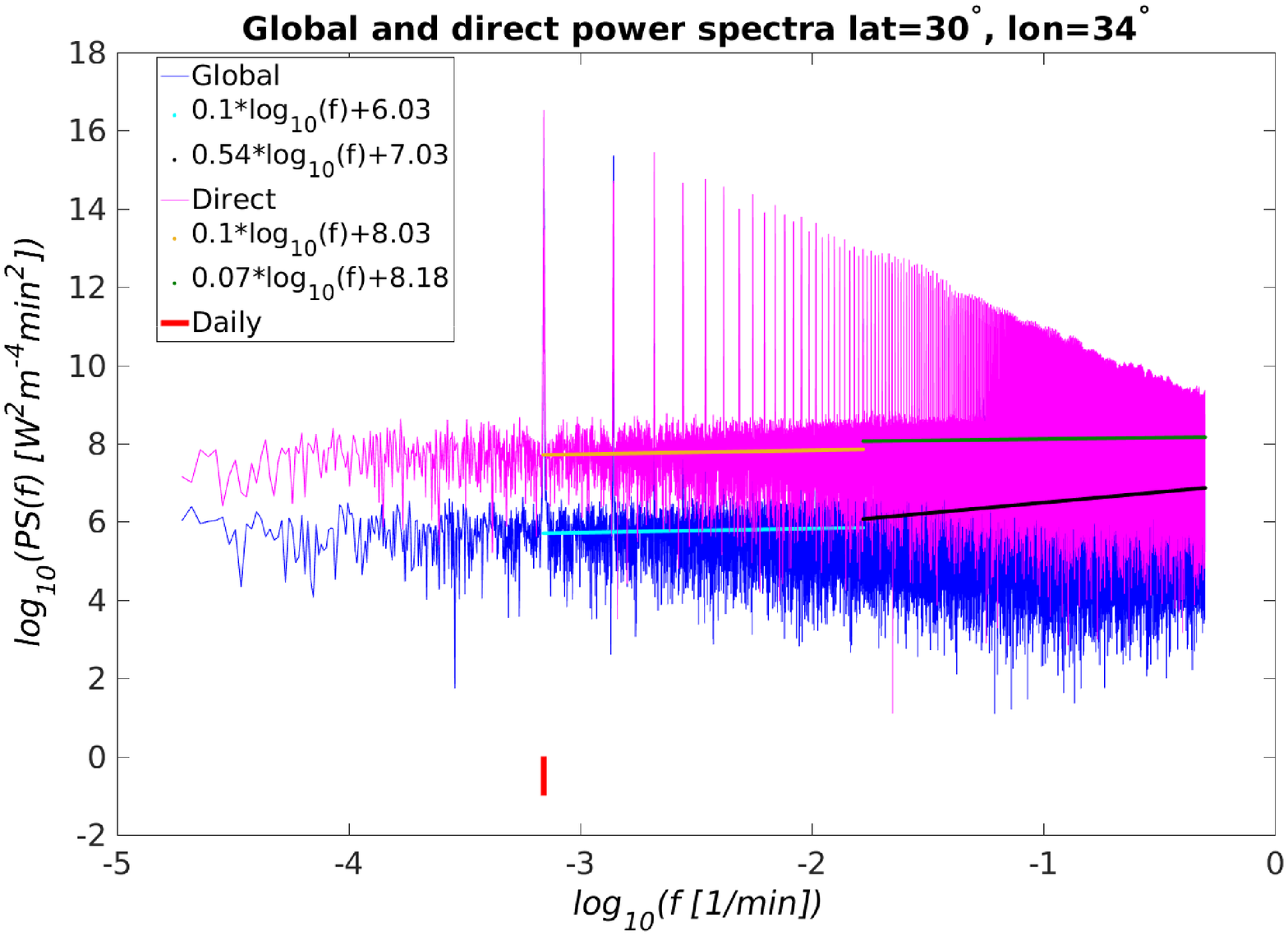}
  \includegraphics[width=0.45\columnwidth]{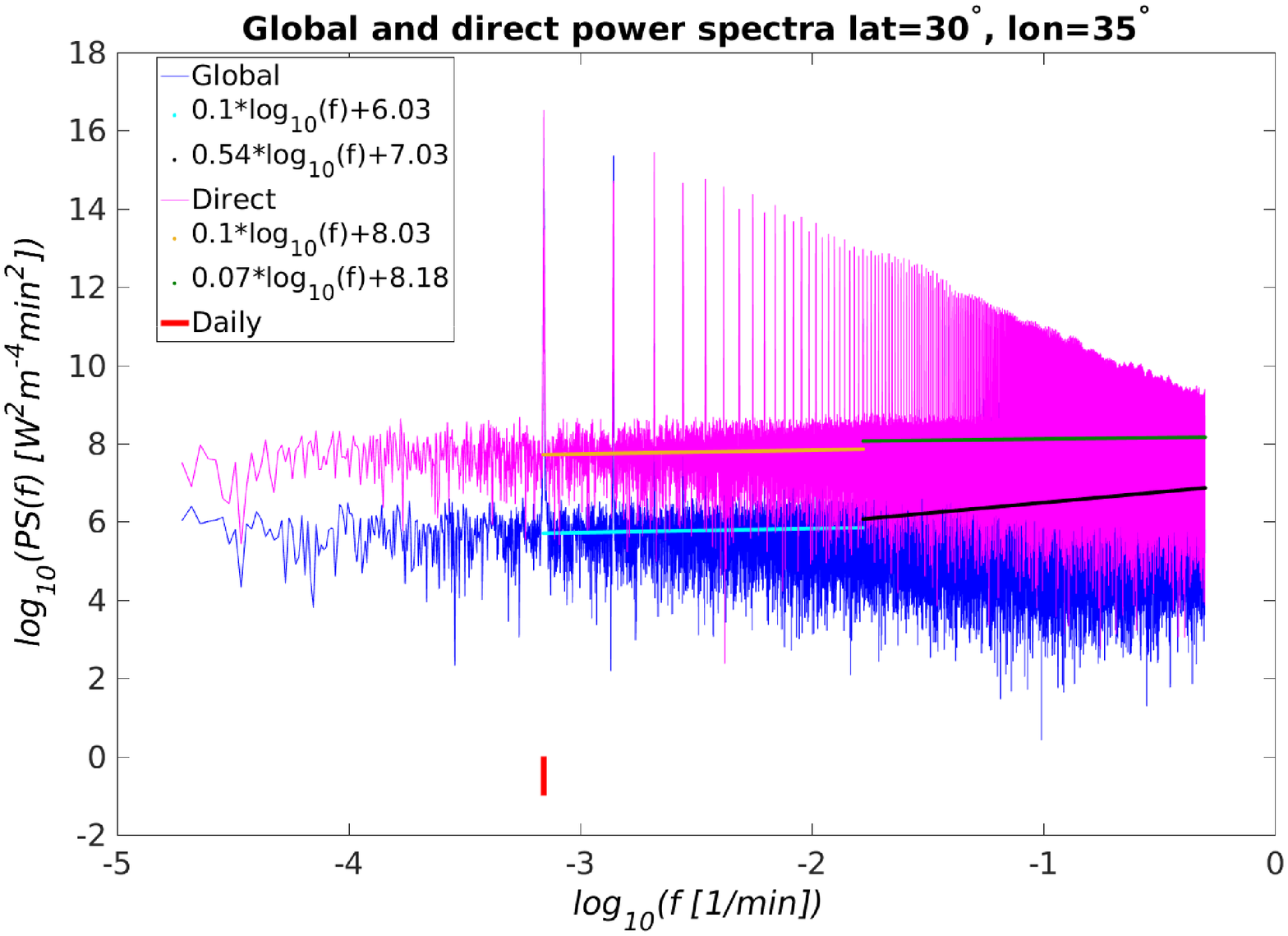}
  \includegraphics[width=0.45\columnwidth]{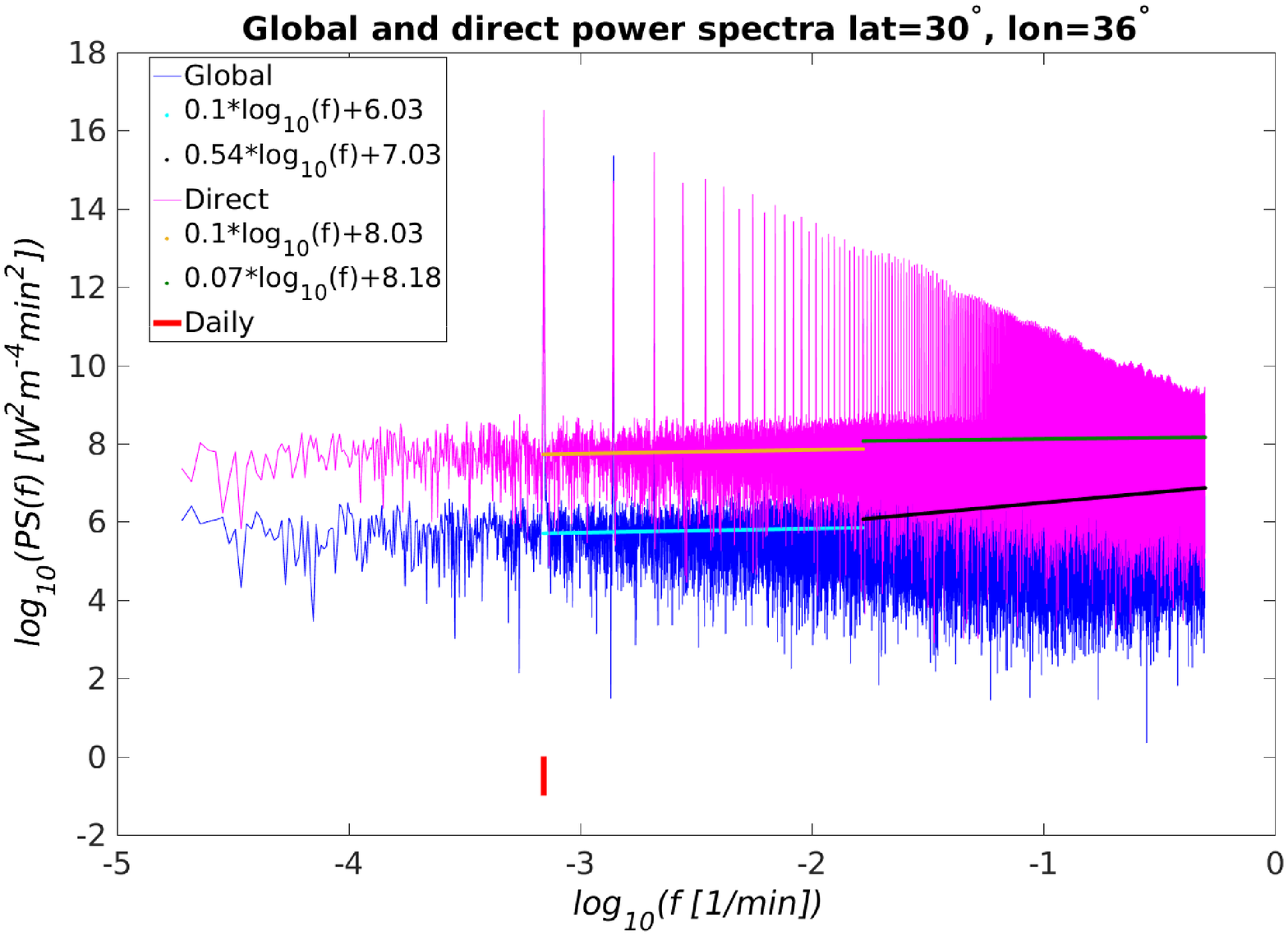}
  \includegraphics[width=0.45\columnwidth]{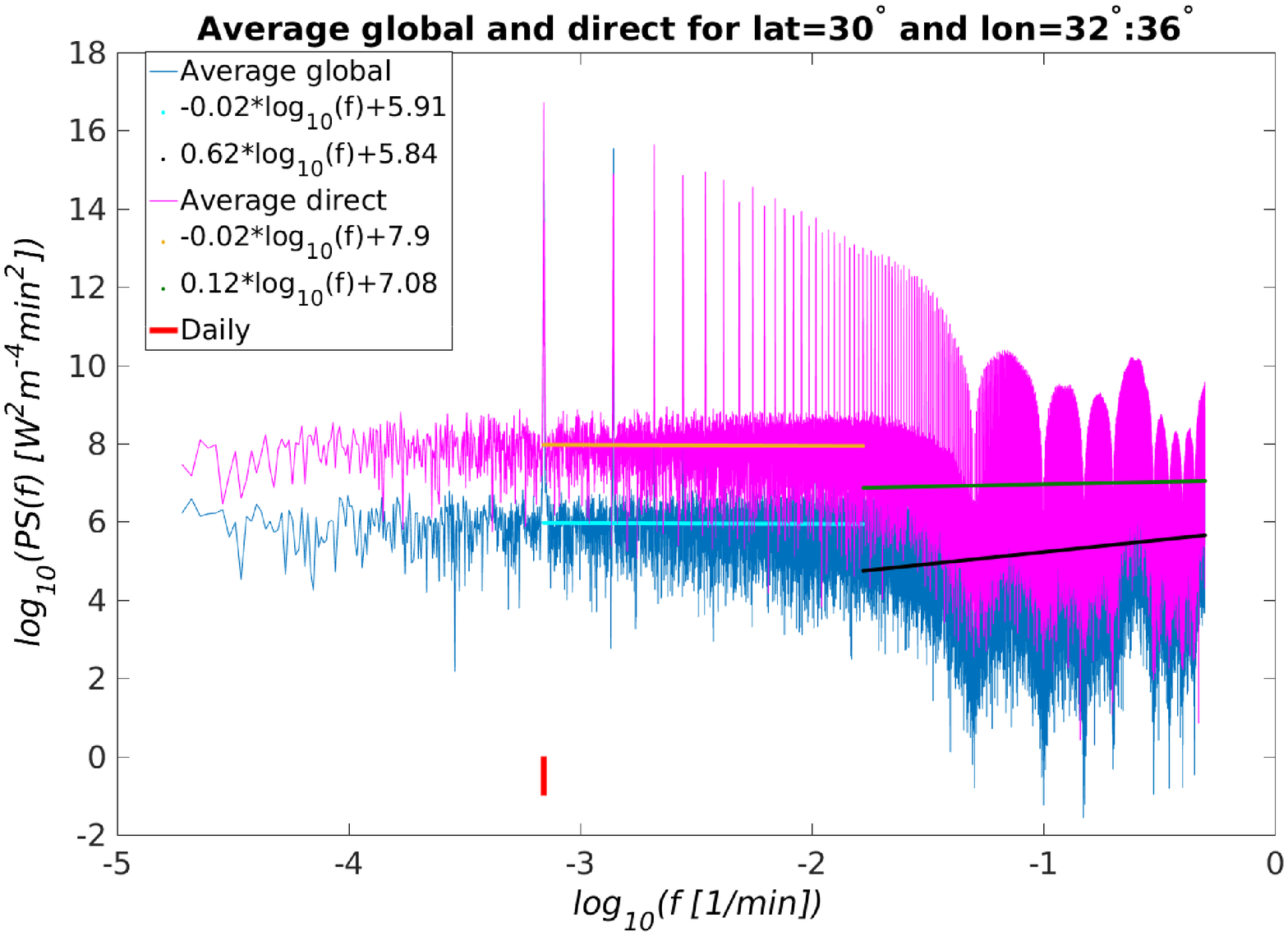}
  \caption{The power spectra of the global and direct components at five different locations with the same latitude, $30^\circ$, and longitudes varying by one degree from $32^\circ$--$36^\circ$.
  We present the power spectra of each component and the power-law fit at the two different frequency ranges. The lowest right panel presents the same information for the averaged, over these five locations, radiation.}
  \label{fig:pslat30}
\end{figure*}

\end{document}